\documentclass[fleqn,usenatbib]{mnras}

\usepackage{newtxtext,newtxmath}
\usepackage[T1]{fontenc}
\usepackage{ae,aecompl}
\usepackage{enumitem}

\usepackage{graphicx}	% Including figure files
\usepackage{amsmath}	% Advanced maths commands
\usepackage{amssymb}	% Extra maths symbols
\usepackage{float} 
\usepackage[dvipsnames]{xcolor}
\usepackage{xspace}
\usepackage[para]{threeparttable} % tables with footnotes

\newcommand{\Msun}{\,M$_\odot$}

\title[\textsc{METISSE}]{The fates of massive stars: exploring uncertainties in stellar evolution with \textsc{METISSE}}

\author[P. Agrawal et al.]
{Poojan Agrawal$^{1,2}$\thanks{E-mail: pagrawal@astro.swin.edu.au}, 
Jarrod Hurley$^{1,2}$,
Simon Stevenson$^{1,2}$, 
Dorottya Sz\'ecsi$^{3}$, 
\newauthor
Chris Flynn$^{1}$
\\
$^{1}$Centre for Astrophysics and Supercomputing, Swinburne University of Technology, Hawthorn, VIC 3122, Australia.\\
$^{2}$OzGrav:  The ARC Centre of Excellence for Gravitational Wave Discovery\\
$^{3}$I. Physikalisches Institut, Universit\"at zu K\"oln, Z\"ulpicher-Str. 77, D-50937 Cologne, Germany
}

\date{Accepted XXX. Received YYY; in original form ZZZ}

\pubyear{2020}

\begin{document}
\label{firstpage}
\pagerange{\pageref{firstpage}--\pageref{lastpage}}
\maketitle

\begin{abstract}

In the era of advanced electromagnetic and gravitational wave detectors, it has become increasingly important to effectively combine and study the impact of stellar evolution on binaries and dynamical systems of stars. Systematic studies dedicated to exploring uncertain parameters in stellar evolution are required to account for the recent observations of the stellar populations.
We present a new approach to the commonly used \textsc{Single-Star Evolution (SSE)} fitting formulae, one that is more adaptable: \textsc{Method of Interpolation for Single Star Evolution (METISSE)}.
It makes use of interpolation between sets of pre-computed stellar tracks to approximate evolution parameters for a population of stars.
We have used \textsc{METISSE} with detailed stellar tracks computed by the \textsc{Modules for Experiments in Stellar Astrophysics (MESA)}, the \textsc{Bonn Evolutionary Code (BEC)} and the Cambridge \textsc{STARS} code.
\textsc{METISSE} better reproduces stellar tracks computed using the STARS code compared to \textsc{SSE}, and is on average three times faster.
Using stellar tracks computed with \textsc{MESA} and \textsc{BEC}, we apply \textsc{METISSE} to explore the differences in the remnant masses, the maximum radial expansion, and the main-sequence lifetime of massive stars. We find that different physical ingredients used in the evolution of stars, such as the treatment of radiation dominated envelopes, can impact their evolutionary outcome. For stars in the mass range 9 to 100\Msun{}, the predictions of remnant masses can vary by up to 20\Msun{}, while the maximum radial expansion achieved by a star can differ by an order of magnitude between different stellar models. 

\end{abstract}

\begin{keywords}
stars: evolution -- methods: numerical -- stars: massive -- stars: black holes -- gravitational waves -- stars: winds, outflows
\end{keywords}

%%%%%%%%%%%%%%%%%%%%%%%%%%%%%%%%%%%%%%%%%%%%%%%%%%

%%%%%%%%%%%%%%%%% BODY OF PAPER %%%%%%%%%%%%%%%%%%

\section{Introduction}
\label{sec:introduction}

Modelling the integrated properties of stellar systems such as galaxies or star clusters requires the use of population synthesis codes which can simulate a large number of stars (a population) and the myriad interactions between them. In order to produce realistic models of such systems which can be compared to modern observations \citep*[e.g.][]{2008LNP...760..347M}, it is important to include an up to date treatment of stellar evolution. 

Stellar evolution is typically modelled using a one-dimensional (1D) stellar structure and evolution code, which we refer to as a `detailed stellar evolution code'. Such codes solve the differential equations of stellar structure (namely for mass, momentum and energy conservation, energy generation and transport) within the star, at different points in time to compute a sequence of stellar structure models. Detailed evolution codes are a recommended way to evaluate both the structure and the evolution of stars but running them for a population of stars can be computationally expensive and time consuming. 

With the advent of high-performance computers and parallel programming methods, detailed evolution codes are being used in combination with stellar dynamics and population synthesis codes e.g. \citet*{2009PASA...26...92C} and Astrophysical Multipurpose Software Environment \citep[AMUSE;][]{2009NewA...14..369P,2013CoPhC.184..456P,2013A&A...557A..84P}. However, detailed stellar evolution codes can break down at times owing to numerical difficulties which can impede the progress of the overlying simulation \citep*{2008LNP...760.....A}. Physical processes such as convection and rotation become important in massive stars and require sophisticated modelling methods with higher temporal and spatial resolution, increasing the computational cost and the potential for numerical issues to develop. User intervention and expertise is often required to push detailed codes past failure points. The data from these simulations also need to be manually checked for any non-physical results which would arise from erroneous numerical evolution of a model star. 
 
While there are considerable differences in the evolutionary tracks for stars of various masses and metallicities, if the step in mass and metallicity is small, the changes are usually smooth enough to parameterize. Furthermore, for most population synthesis requirements only the global parameters of the stars such as mass, radius, and luminosity are needed. Similarities between the stellar tracks can be exploited and the output of a detailed code for a few stars can be parameterized in the form of formulae \citep{1996IAUS..174..213E}. These formulae can then be used to calculate evolution properties for a large number of stars.

The earliest attempts to include the effects of stellar evolution in the study of star clusters were made by \citet{1970CeMec...2..353W}, \citet{1987MNRAS.224..193T} and \citet{1990ApJ...351..121C}. The authors employed simple schemes for stellar lifetimes and only accounted for the mass lost in the form of planetary nebulae or during supernovae events. A more accurate method was developed by \citet*{Hurley2000} in the form of the \textsc{Single Star Evolution (SSE)} package obtained using polynomial fits to the set of stellar tracks by \citet{1998MNRAS.298..525P}. It was an expansion of the work by \citet*{1989ApJ...347..998E} along the lines of \citet{1997MNRAS.291..732T}. The \textsc{SSE} package employs fitting formulae and analytical expressions for the underlying physics to describe quantities such as the radius and luminosity of a star given its mass, metallicity and age. Fitting formulae have been a popular choice for population synthesis codes because the resulting algorithms are computationally inexpensive, fast and robust.
 
Two decades later, ground-based telescopes such as the \textit{Very Large Telescope} \citep{1998Sci...280.1520S,2009ASSP....9.....M} and \textit{Keck} \citep{2018SPIE10702E..07K} have been observing fainter and rarer stars while the \textit{Hubble Space Telescope} \citep{1991epu..conf..371P,1994ASSL..187...87S}, \textit{Chandra X-ray Observatory} \citep{2019A&G....60f6.19W} and \textit{Gaia} \citep{2012Ap&SS.341...31D, 2019arXiv191207659E} have monitored complex stellar phenomena from space. Furthermore, interferometers such as the \textit{Very Large Array} and the \textit{Atacama Large Millimeter Array} have helped us probe the formation and afterlives of stars through radio observations \citep{2019PASP..131a6001M}.
Advances in multi-messenger astronomy have also provided us with unprecedented data with which we can better understand the universe. The \textit{IceCube Neutrino Observatory} \citep{2020NIMPA.95261650W} is detecting high energy neutrinos from stellar outbursts, while the \textit{Advanced Laser Interferometer Gravitational-wave Observatory} \citep[aLIGO;][]{2015CQGra..32k5012A} and  \textit{Advanced Virgo} \citep{2015CQGra..32b4001A} detectors continue to report gravitational-wave observations from the merging of compact binaries \citep{2016PhRvL.116f1102A,2017PhRvL.119p1101A,2018arXiv181112907T,2020arXiv200408342T}. 

Together with the advances in our observing capabilities the development of sophisticated numerical techniques in programming and newer input data in the form of opacity tables and nuclear reaction rates has led to the development of modern and improved stellar structure and evolution codes with updated physics \citep{2019ApJS..243...10P}. Thus, there is a pressing need to update the fitting formulae used in \textsc{SSE} using the data from up-to-date stellar evolution tracks. 

Re-calculating the fitting formulae from a new set of stellar tracks is a non-trivial task \citep{2009PASA...26...92C}. \citet{2019arXiv190606641T} recently performed an update of the \textsc{SSE} formulae for metal-poor massive stars. However, even with the updated fitting formulae, this only covers a particular subset of the parameter space and the user is still limited to results from a single set of evolutionary tracks. There is thus a need for a more flexible method which is also fast, robust and can easily make use of different stellar evolution tracks. 

Interpolation between a set of pre-calculated evolutionary tracks provides a promising alternative. This method employs tabulated data from 1D stellar evolution codes to estimate stellar parameters for a desired star. Unlike fitting formulae, stellar parameters from the given set of detailed tracks are calculated in real time with this method. Hence, one just needs to change the input stellar tracks to generate a new set of stellar parameters.

Although interpolation between stellar tracks has been extensively used to construct stellar isochrones \citep[e.g.][]{1992A&AS...96..269S, 2001ApJ...556..322B}, the memory requirement for storing and loading the tracks made it difficult for computationally expensive codes involving stellar dynamics to make use of interpolation in the past. 
With modern computers, computer memory is readily available and recently, the codes \textsc{SEVN} \citep*{2015MNRAS.451.4086S,2017MNRAS.470.4739S} and \textsc{ComBinE} \citep{2018MNRAS.481.1908K} have employed the method of interpolation over a range of stellar parameters to study the properties of gravitational wave progenitors. Presently, interpolation offers the most viable option for an efficient, robust and flexible approach.

In this paper, we present results from our newly developed synthetic stellar evolution code \textsc{METhod of Interpolation for Single Star Evolution (METISSE}). It uses interpolation to approximate the properties of a star of given mass and metallicity at any age. It is a modern Fortran code and can serve as an alternative to \textsc{SSE} fitting formulae in stellar dynamics and population synthesis codes. It relies on the concept of Equivalent Evolutionary Phases (EEPs) and can make use of stellar tracks from a variety of stellar evolution codes. In this work, we have used sets of stellar tracks computed using the Cambridge STARS code, \textsc{Modules for Experiments in Stellar Astrophysics (MESA)} and the \textsc{Bonn Evolutionary Code (BEC)} as input to \textsc{METISSE}. Using the \textsc{MESA} and \textsc{BEC} tracks in \textsc{METISSE}, we predict stellar parameters such as the maximal extent of the radius or the remnant mass for massive stars and compare the results in terms of their physical ingredients. We thus demonstrate the usefulness of \textsc{METISSE} in systematic studies dedicated to exploring how uncertain parameters in stellar evolution effect the properties of binary populations and dynamical systems of stars.

This paper is organized as follows. We provide an overview of evolutionary tracks for different stars and the concept of EEPs in Section~\ref{sec:stellar_life}.
We describe the construction of \textsc{METISSE} as a standalone stellar evolution code in Section~\ref{sec:METISSE}. In Section~\ref{sec:stellar_models} we introduce the three sets of stellar models that we have used to show \textsc{METISSE}'s capabilities. We validate results obtained with \textsc{METISSE} by comparing to \textsc{SSE} in Section~\ref{sec:test_METISSE}. In Section~\ref{sec:compare_stellar_models}, we present results from \textsc{METISSE} using stellar tracks computed with \textsc{MESA} and \textsc{BEC} as input.
We mention the key differences between these tracks and their implications in Section~\ref{sec:discussion_of_results}. We discuss caveats and potential future work in Section~\ref{sec:caveats_and_future_work} and conclude the paper in Section~\ref{sec:conclusions}.

\section{Stellar life and EEPs}
\label{sec:stellar_life}

\begin{figure}
	\includegraphics[width=\columnwidth]{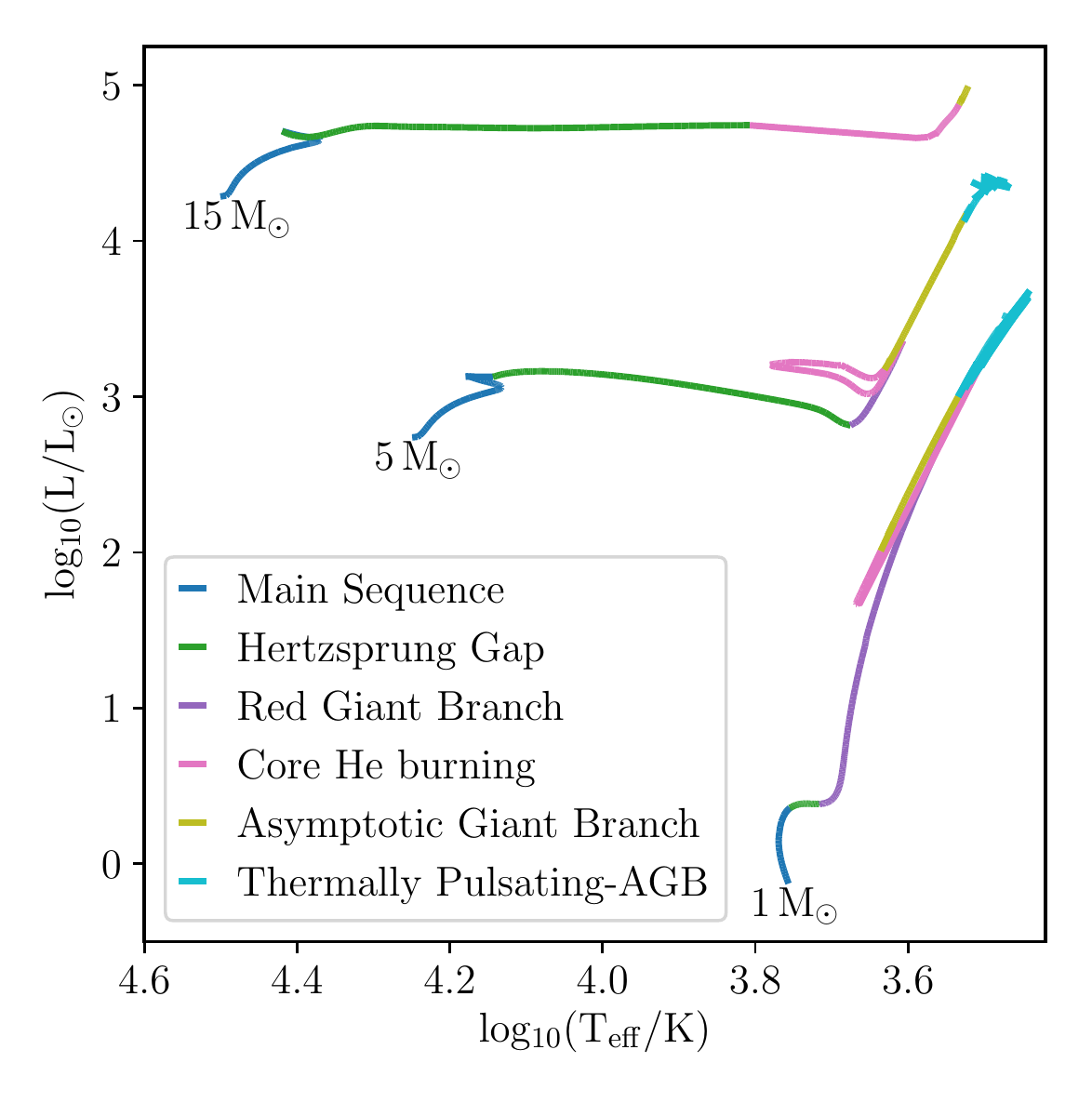}
    \caption{Hertzsprung--Russell (HR) diagram showing evolutionary tracks for stars of mass 1, 5 and 15\Msun{} at a metallicity of $Z = 0.0142$. Different evolutionary phases are highlighted along each track. The post-asymptotic giant branch phase has not been plotted for clarity.}
    \label{fig:phases}
\end{figure}

Stars have varied lives depending on their mass and chemical composition.
Owing to the differences in their evolution, stars experience different evolutionary phases and trace different paths on a Hertzsprung--Russell (HR) diagram \citep[see, e.g.,][for an in-depth discussion of the evolution of stars]{1968pss..book.....C, 1990sse..book.....K}. Stellar tracks highlighting different evolutionary phases are shown in Fig.~\ref{fig:phases}. 

A low-mass star like our Sun (1\Msun{}) burns hydrogen (H) via the proton-proton chains in a radiative core. This causes the surface temperature and the luminosity to increase moderately while the star is on the main sequence (MS). At the end of MS, the core is not hot enough to ignite helium (He) and contracts, becoming degenerate at some point on the Hertzsprung Gap (HG). The envelope, however, cools and expands as the star ascends the Red Giant Branch (RGB). The H burning in the shell surrounding the core adds to the core mass until it becomes hot and massive enough to ignite He off-center in a thermonuclear run-away (He flash). The star descends the giant branch as the core expands due to a decrease in the hydrostatic pressure and burns He in the core while on the horizontal branch. It ascends the Asymptotic Giant Branch (AGB) at the end of core He burning, and then transitions to the Thermally Pulsating-AGB (TPAGB) where it eventually loses its envelope to become a white dwarf (WD).

An intermediate-mass (e.g. 5\Msun{}) star, on the other hand, burns H via the Carbon-Nitrogen-Oxygen cycle in a convective core. The effective temperature of the star decreases during the main-sequence evolution, making it move redwards on the HR diagram. Mixing by convection causes the sudden depletion of H in the region surrounding the core. In the absence of nuclear energy generation in the core, the star contracts on a thermal time-scale, producing the hook-like feature on the HR diagram seen at the end of the MS. H-shell burning ensues as the star ascends the giant branch. He ignition happens at the tip of the RGB, in semi-degenerate conditions without a flash, and the star burns He in a blue loop, ascending the AGB at the end of He burning and ending life most likely as a carbon-oxygen white dwarf.

Massive stars (e.g. 15\Msun{}) behave similarly to intermediate-mass stars during the main-sequence phase. Their core becomes hot enough to ignite He on the HG, close to the end of the MS so these stars do not become red giants as low and intermediate-mass stars do. Instead, they continue fusing elements in the core while rapid shell burning adds to the core mass and causes the envelope to slowly expand, thereby making the star a red supergiant (RSG). Finally, with the formation of an iron core, the star runs out of fuel and ends its life in a supernova (SN) explosion.
 
Modelling stars through different evolutionary phases using detailed evolution codes typically requires numerous and unequal steps in time. Using the output of a detailed code directly to create an interpolated new track can thus be inefficient and even inaccurate. A track obtained by sequentially interpolating between the same numbered lines in neighbouring mass tracks, might not represent the actual evolution of the star (the evolution we would obtain by simulating the star through detailed codes). Using time as a parameter for interpolation would also not serve the purpose as the associated timescales can again be different for different mass stars. For example, it takes about 10~billion years for a 1\Msun{} star to complete H burning in its core while a 15\Msun{} star can complete all the fusion reactions and form a remnant in just a few million years.

Utilizing evolutionary features such as the depletion of the central hydrogen mass fraction to a certain value along stellar tracks \citep[similar to][]{1970ApJ...159..895S} provides more accurate ground for comparison. These features mark the boundary of evolutionary phases in a stellar track and divide the track into what are known as Equivalent Evolutionary Phases \citep[EEPs;][]{1976PhDT.........8P,2001ApJ...556..322B}. For different stellar tracks, EEPs are readily identifiable by a set of physical conditions. The portion of an evolutionary track between each EEP is further subdivided into an equally spaced set of points. The final product is an EEP-track containing stellar parameters at a fixed number of points. Depending on how many phases a particular track has, the total number of points on an EEP-track can vary. A new track can be generated by interpolating between corresponding points of the neighboring mass tracks. 

In the remainder of the paper we use the term `stellar model' to mean the same as the sequence of stellar models or a stellar track while the term `set of stellar models' or `set of stellar tracks' means evolutionary tracks of stars with different initial masses but the same metallicity.

\begin{table*}
	\caption{\textsc{SSE} phases with the EEP name used by \textsc{METISSE} to identify the start of each phase and the corresponding EEP number.} 
	\label{tab:EEP_table}
	\begin{threeparttable}
	\begin{tabular}{llll} 
        \hline
        No. & Stellar Phase & EEP name & EEP value$^{\rm a}$\\
        \hline
        0 & Main Sequence (MS) ${\rm M<=0.7\,M\sun{}}$ & Zero-Age Main Sequence (ZAMS) & 202\\
        1 & Main Sequence (MS) ${\rm M>0.7\,M\sun{}}$ & Zero-Age Main Sequence (ZAMS) & 202\\
        2 & Hertzsprung Gap (HG) & Terminal-Age Main Sequence (TAMS) & 454\\
        3 & First Giant Branch (GB) & Base of the Giant Branch (BGB) & $^{\rm b}$\\
        4 & Core Helium Burning (cHeB) & core He Ignition (cHeI) & 605\\
        5 & Early Asymptotic Giant Branch (EAGB) &  Terminal-Age core He Burning (TAcHeB) & 707\\
        6 & Thermally Pulsating AGB (TPAGB) & TPAGB & 808\\
        7 & Naked Helium Star MS (HeMS) & None & --\\
        8 & Naked Helium Star HG (HeHG) & None & --\\
        9 & Naked Helium Star Giant Branch (HeGB) & None& --\\
        10 & Helium White Dwarf (HeWD) & None& --\\
        11 & Carbon-Oxygen White Dwarf (COWD) & None& --\\
        12 & Oxygen-Neon White Dwarf (ONeWD) & None& --\\
        13 & Neutron Star (NS) & None& --\\
        14 & Black Hole (BH) & None& --\\
        15 & Massless remnant & None& --\\
    	\hline
	\end{tabular}
	
	\begin{tablenotes}
	\textbf{Notes.}\\
	    \item[a, b] The EEP values here denote the default in \textsc{METISSE} and correspond to the location of primary EEPs from \citet{2016ApJ...823..102C}, except for the BGB EEP which is identified separately for each track. For different stellar models, the value of these EEPs (including the BGB EEP) can be redefined by the user.\\
        For phases 7-15, see section~\ref{subsec:stellar_phases} for how these are calculated. 
      \end{tablenotes}
    \end{threeparttable}
\end{table*}

\section{\textsc{METISSE}}
\label{sec:METISSE}

\textsc{METhod of Interpolation for Single Star Evolution (METISSE)} is a synthetic stellar evolution code which uses interpolation to compute evolutionary tracks for many stars using tracks for a finite set of stars. The tracks for input are evolved using detailed stellar evolution codes and should be converted to EEP form for use in \textsc{METISSE}. The EEPs can be identified with programs like ISO \citep[e.g.][]{2016ApJS..222....8D} or by direct inspection \citep[e.g.][]{szcsi2020bonn}. 
Given a set of EEP-tracks, a schematic of how \textsc{METISSE} calculates the properties of a star within the input mass range is described next.

\subsection{Interpolation scheme}
\label{subsec:interpolation_scheme}

The mass interpolation routine used in this work is adapted from the ISO code \citep{2016ApJS..222....8D}. For a particular value of metallicity, first the corresponding EEP-tracks are read by \textsc{METISSE}. Next, the tracks with initial masses that immediately envelop the input mass are located from the given set. A new track is interpolated by the method of monotonic interpolation with a  piece-wise cubic function \citep{1990A&A...239..443S}. No interpolation occurs at this stage if the track for the mass in question is already present in the set (up to some tolerance defined by the user).

Depending on the metallicity, stars greater than a certain mass do not undergo some evolutionary phases (e.g. the red-giant branch). Interpolation between tracks where some undergo a certain phase and others do not, can result in an incorrect new track. To handle this we identify certain critical mass tracks in the set of EEP tracks for a given metallicity. 
Both the search and the interpolation method change if the input mass falls near a critical mass, such as the mass above (or below) which stars do (or do not) ignite He on the HG. In this case, the track is either linearly interpolated or extrapolated if necessary. In Section~\ref{subsec:Z_par} we provide details on how these critical masses are identified. 

The mass interpolated track, however, contains stellar parameters for a set of ages. These generally differ from the age at which evolution parameters are required by a population synthesis code. So another interpolation is performed in age within the newly interpolated track to return stellar parameters at any given time. 

\subsection{Stellar phases}
\label{subsec:stellar_phases}

From an input set of models, \textsc{METISSE} determines the location of certain major EEPs to assign stellar evolution phases similarly to \textsc{SSE} \citep{Hurley2000} to the interpolated tracks. The key EEPs and the corresponding \textsc{SSE} phases are listed in Table~\ref{tab:EEP_table}.
To ensure that the interpolation occurs between equivalent evolutionary phases for each star, each stellar phase should occur at the same EEP value and hence at the same line number across the input stellar tracks. For evolutionary phases that do not occur in all evolutionary tracks, the EEP value is treated as a continuation of the preceding phase. For example, the base of the giant branch (BGB) may be missing for massive stars, so the BGB EEP there is treated as a part of the HG.

As outlined in Section~\ref{sec:stellar_life}, low and intermediate-mass stars enter a remnant phase after losing their envelope on the AGB while high-mass stars fuse elements all the way until iron in their core before becoming a remnant. However, modelling the evolutionary phases beyond carbon burning is numerically difficult and the phases themselves are short lived, hardly contributing to the overall evolution of the stars. Hence, we assume that the star has reached the end of its life when it either reaches the end of the detailed track during the AGB phase or when the carbon-oxygen core mass exceeds the maximum allowed core mass \cite[c.f. equation 75 of][]{Hurley2000}:

\begin{equation}
M_{\mathrm{c}, \mathrm{SN}}=\max \left(M_{\mathrm{ch}}, 0.773 M_{\mathrm{c}, \mathrm{BAGB}}-0.35\right),
\label{eqn:mcmax}
\end{equation}
where $M_{\mathrm{ch}}$ denotes the Chandrasekhar mass and $M_{\mathrm{c,BAGB}}$ is the core mass at the start of the AGB phase of the star. The stellar parameters at this stage are used to determine the type and the properties of the remnant that the star would form. Corresponding parameters are calculated using the methods described in Section~\ref{subsec:stellar_remnants}.

At each step, we also check if the star has lost its hydrogen envelope. For massive single stars, this can occur during late evolutionary stages. For low-mass stars this can only occur in binary systems where mass transfer prematurely removes the envelope of the donor star. 
The evolution of such stripped (naked helium) stars is different compared to other stars and helium star models are needed to follow their subsequent evolution \citep{2002PASA...19..233P, 2019ApJ...878...49W, 2020A&A...637A...6L}.
Currently in \textsc{METISSE} we revert to using the fitting formulae outlined by \citet{Hurley2000} for evolving stars after they lose their envelope. 
In the future, we will make use of helium star model data in \textsc{METISSE} to treat the evolution of naked helium star phases by interpolating in a set of helium star models in \textsc{METISSE} \citep[as in][]{2019MNRAS.485..889S}.

\section{Stellar models}
\label{sec:stellar_models}

In order to interpolate a stellar track of a given mass and metallicity, \textsc{METISSE} requires a set of EEP-tracks of the same metallicity. These are calculated with detailed evolutionary codes. In this paper, we make use of stellar models calculated using three different detailed stellar evolution codes. Below we describe these models and how they are converted to EEP form for application in \textsc{METISSE}. Additional details about these models are discussed in Section~\ref{sec:discussion_of_results}.

\subsection{POLS98 models}
\label{subsec:pols98_stellar_models}

The POLS98 models were used for computing the original \textsc{SSE} fitting formulae by \citet{Hurley2000} and were evolved by \citet{1998MNRAS.298..525P} using an updated version of the stellar evolution code STARS \citep{1971MNRAS.151..351E}. 
The stellar models cover metallicities between $Z = 0.0001$ and $Z = 0.03$. There are about 25 tracks between 0.5 and 50\,M\sun{} for each metallicity.
Depending on their initial mass, these tracks have been computed from the ZAMS to different end points. The evolution of massive stars was computed until central carbon ignition. For stars with initial mass less than 1\Msun{}, tracks are complete up to the occurrence of the degenerate helium flash while for intermediate-mass stars the evolution ends at the start of the first thermal pulse on the asymptotic giant branch. 

In this paper, we use the sets of tracks labelled as the OVS tracks by \citet{1998MNRAS.298..525P}. The tracks were computed with enhanced mixing (described in Section~\ref{subsec:disc_conv}) and assume no mass loss due to stellar winds. 
For application in \textsc{METISSE}, the tracks are converted into the EEP-format by use of critical turning points defined in Table 2 of \citet{1998MNRAS.298..525P} and a weighted metric function from \citet{2016ApJS..222....8D}.

\subsection{\textsc{MESA} models}
\label{subsec:MESA_stellar_models}

Modules for Experiments in Stellar Astrophysics \citep[MESA;][]{2019ApJS..243...10P} is a modern, open-source stellar evolution code. In order to test \textsc{METISSE}, we have used \textsc{MESA} version 11701 to compute a set of stellar tracks for metallicity $Z = 0.00142$. The set consists of 25 tracks of non-rotating single stars between 9 and 200\Msun{}. The tracks have been computed from the pre-main sequence until carbon depletion ($X_c \leq 10^{-4}$) in the core, although for the purposes of testing \textsc{METISSE}, only the phases after the ZAMS are relevant.

We have employed the standard \textsc{MESA} Dutch scheme \citep{2009A&A...497..255G} for stellar wind mass loss. We also include the contribution to mass loss owing to super-Eddington winds in our models (see Section~\ref{subsec:disc_mass_loss_scheme} for details). An extensive nuclear reaction network of 77 elements has also been used to closely follow the evolution of massive stars while other input parameters are the same as given by \citet{2016ApJ...823..102C}. These tracks and more details about them will be published in another paper (Agrawal et al. in preparation). 
Output tracks from \textsc{MESA} have been converted into EEP-format with ISO \citep{2016ApJS..222....8D}.

\subsection{\textsc{BEC} models}
\label{subsec:BEC_stellar_models}

The Bonn Code, which we refer to as `\textsc{BEC}' in this paper, is a detailed stellar evolution code which has been used in the last decades in various science projects (see e.g. \citealp*{Heger:2000a}, \citealp{Petrovic:2005}, \citealp*{Yoon:2006} and references therein). Here we apply a set of models computed with this code and published in the BoOST project \citep[][]{szcsi2020bonn}. These models are slowly rotating (at about $100$\,km s$^{-1}$) and have been computed from the ZAMS until the end of core helium burning.

The BoOST project published stellar models as well as interpolated tracks between these models. Here we have made use of only the former. We use their dwarfA set of models which have a metallicity $Z = 0.00105$. The tracks are optimized for astrophysical applications such as population synthesis and the format of the published models does already fulfil the requirements of the EEP-tracks.

\section{Testing \textsc{METISSE} with POLS models}
\label{sec:test_METISSE}

\begin{figure}
  \includegraphics[width=\columnwidth]{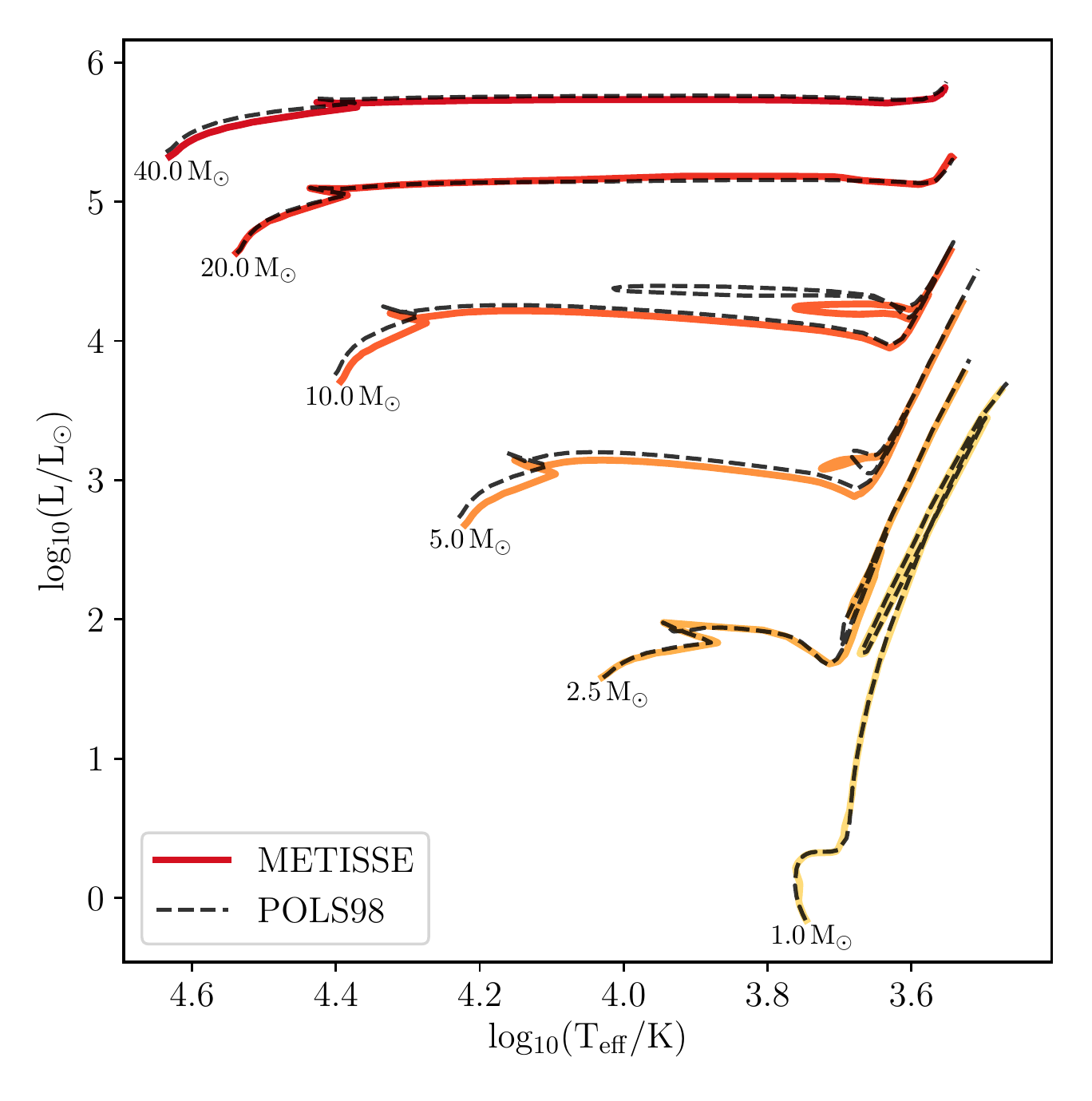}
  \caption{HR diagram showing tracks interpolated by \textsc{METISSE} (solid lines) with detailed tracks by \citet{1998MNRAS.298..525P} and the detailed tracks themselves (dashed lines) for a metallicity of $Z = 0.02$. For each mass, the detailed track was removed from the set before performing interpolation.}
  \label{fig:interpolated}
\end{figure}

The main requirement of a synthetic stellar evolution code such as \textsc{METISSE} is for the interpolated tracks to replicate the underlying detailed evolutionary tracks as closely as possible. In this section we check the accuracy of \textsc{METISSE} by comparing its output with the detailed models and we also compare the results obtained by \textsc{METISSE} with those by \textsc{SSE} \citep{Hurley2000}. 
To make a direct comparison with \textsc{SSE}, the stellar tracks generated with \textsc{METISSE} use the set of detailed tracks by \citet{1998MNRAS.298..525P} as input. 
Because the input models do not include mass loss in stellar winds, all the results shown in this section, with both \textsc{SSE} and \textsc{METISSE}, do not have mass loss enabled either, except during the formation of the remnant (in the form of planetary nebula or supernova ejecta, cf. Sections~\ref{subsec:stellar_remnants} and \ref{subsec:stellar_phases}).

\subsection{Accuracy of interpolated tracks}
\label{subsec:accuracy_of_interpolated_tracks}

To test the quality of tracks computed with \textsc{METISSE}, we interpolated evolutionary tracks for certain initial masses present in the \citet{1998MNRAS.298..525P} set of detailed models. Usually, if an EEP-track is already present in the set of input tracks, \textsc{METISSE} would simply return that track and would not perform an interpolation in mass. Hence, we sequentially removed the detailed track for each input mass from the set before interpolating a new track. The interpolated tracks and the corresponding detailed tracks from \citet{1998MNRAS.298..525P} are shown in Fig.~\ref{fig:interpolated}. 

We find that the tracks interpolated by \textsc{METISSE} are in good agreement with the detailed tracks. To quantify this agreement we calculate the relative difference in luminosity $(L)$ and surface temperature $(T_{\rm eff})$ between detailed and mass interpolated EEP tracks. 
For most evolutionary phases, the average difference between the track interpolated with the \citet{1990A&A...239..443S} scheme and the detailed track is less than 3 per cent for both quantities. For the core helium burning (blue loop) phase the variation in $L$ can be up to 10 per cent.
The greatest dissimilarity occurs if the input mass is close to a critical mass (cf. Section~\ref{subsec:Z_par}).
In Fig.~\ref{fig:interpolated}, the 5\Msun{} track falls near the critical mass above which C ignition can occur non-degenerately in the core while the 10\Msun{} track falls near the critical mass above which He ignition occurs on the HG. Unlike the other tracks, where third order interpolation has been used, these two tracks have been linearly interpolated from their neighbouring tracks and in this case the average difference can be as high as 21 per cent in $L$ and 13 per cent in $T_{\rm eff}$ during the core helium burning phase. 

We note that the quality of interpolation also depends on the density and the completeness of the input tracks (cf. Section~\ref{sec:caveats_and_future_work}).
For a denser grid of stellar models, tracks interpolated by \textsc{METISSE} mimic detailed tracks even more closely.

\subsection{Comparison with \textsc{SSE}}
\label{subsec:compare_SSE}

\begin{figure}
  \includegraphics[width=\columnwidth]{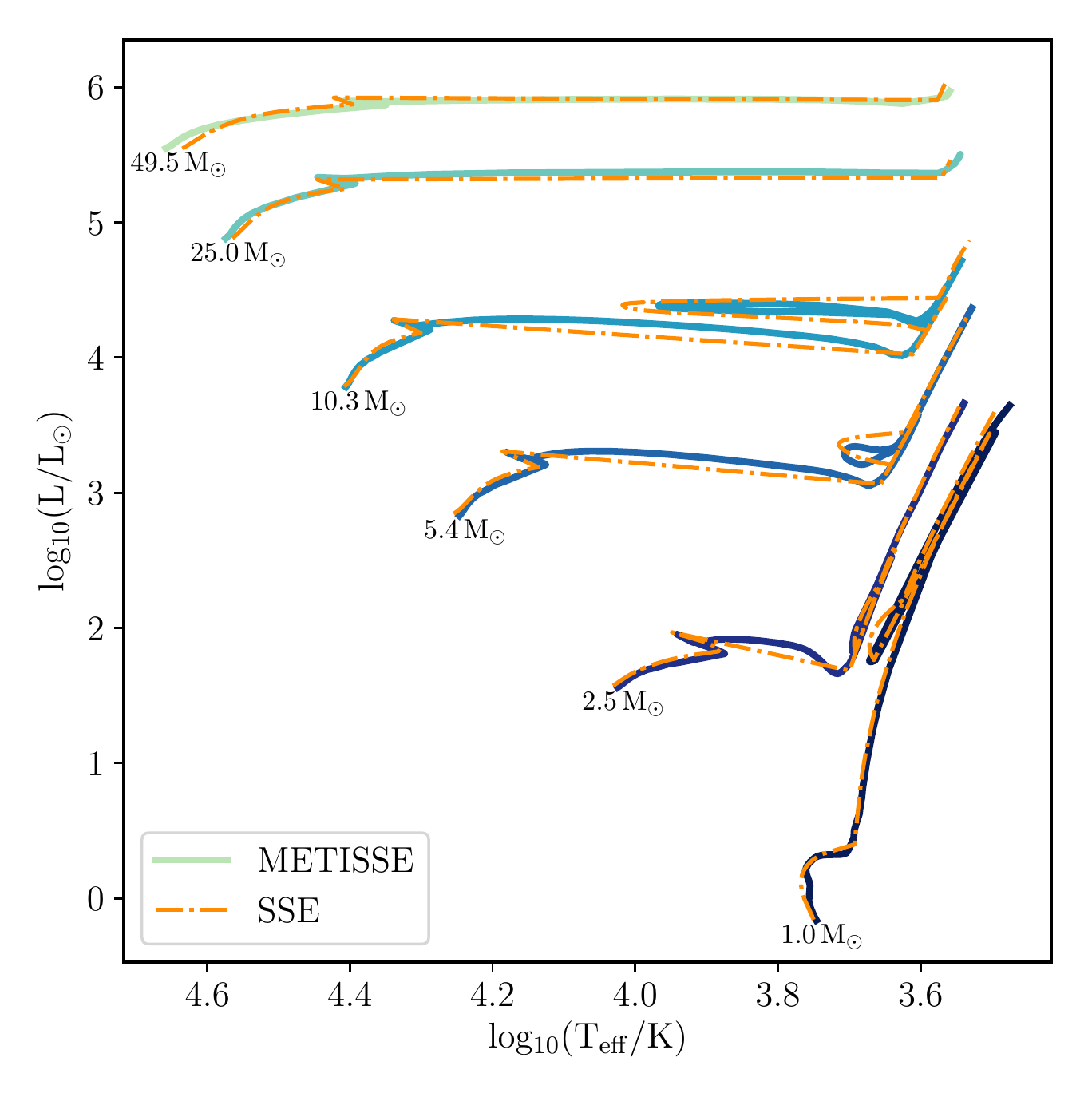}
  \caption{HR diagram comparing tracks interpolated by \textsc{METISSE} (solid lines) with tracks computed by the fitting formulae of \textsc{SSE} (dashed lines) for metallicity $Z = 0.02$. Both methods use detailed tracks by \citet{1998MNRAS.298..525P} as input and assume no mass loss in stellar winds.}
  \label{fig:SSE_METISSE}
\end{figure}

\begin{figure*}
    \begin{tabular}{cc}
    \includegraphics[width=0.9\columnwidth]{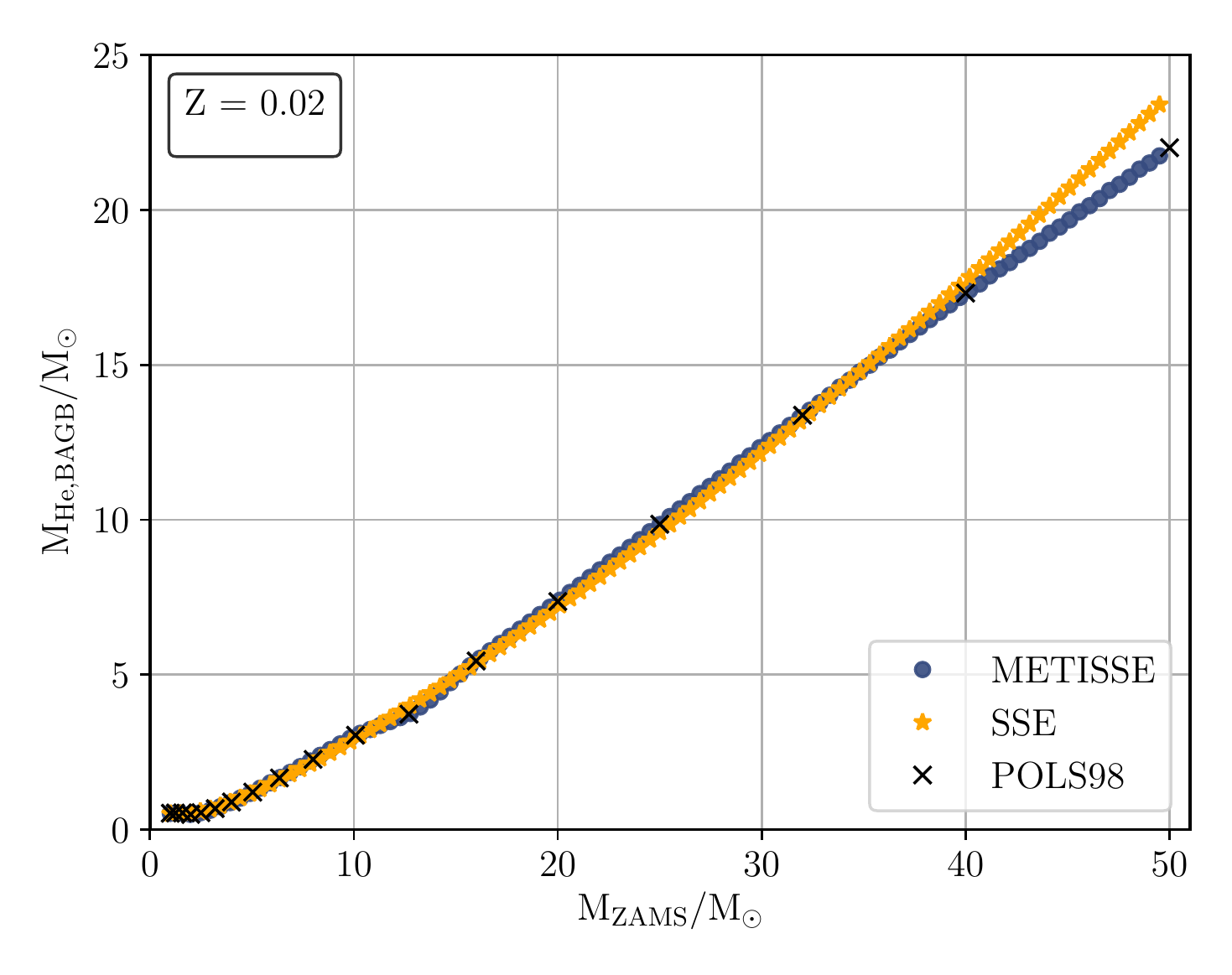}
    & 
    \includegraphics[width=0.9\columnwidth]{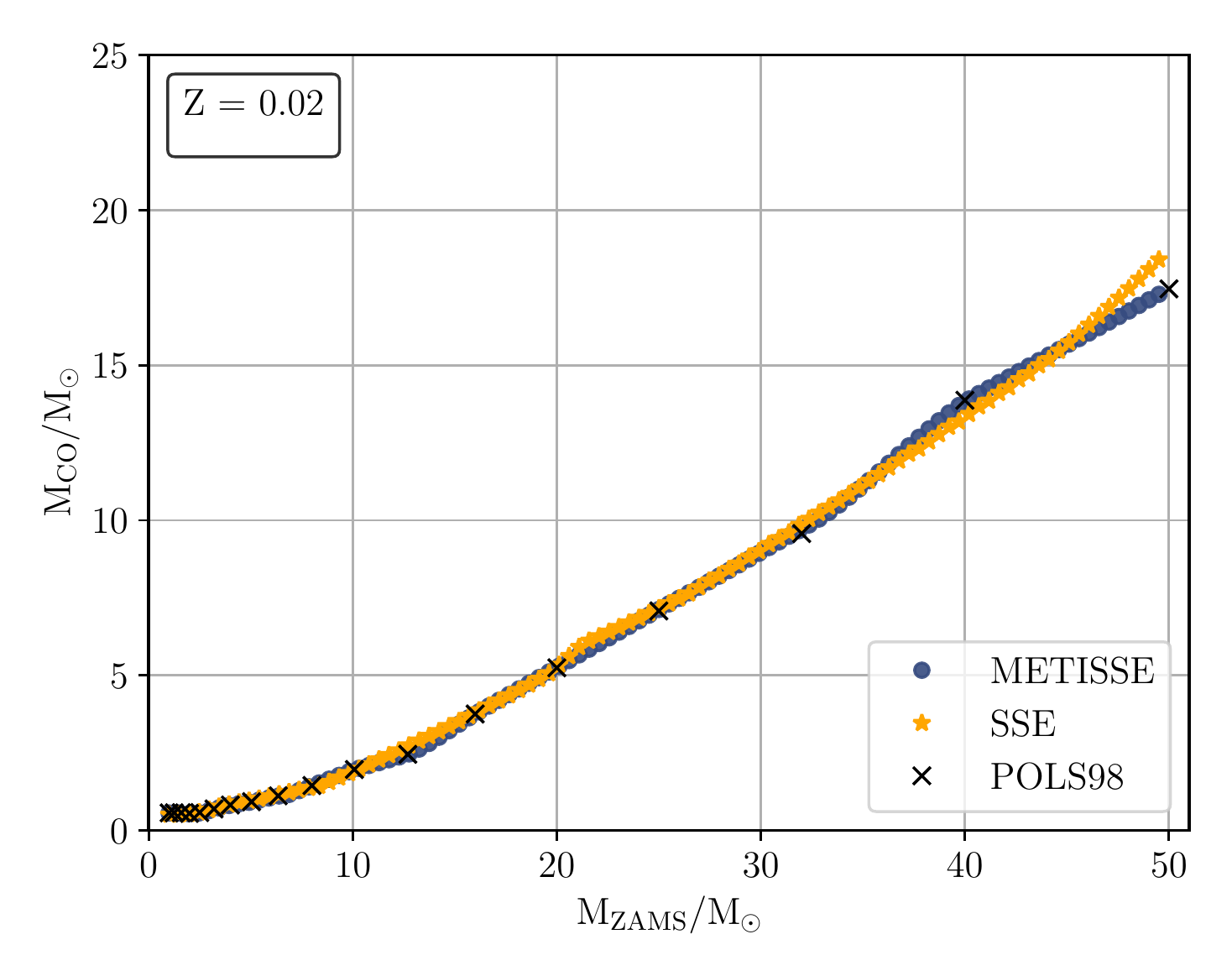}
    \\
    \includegraphics[width=0.9\columnwidth]{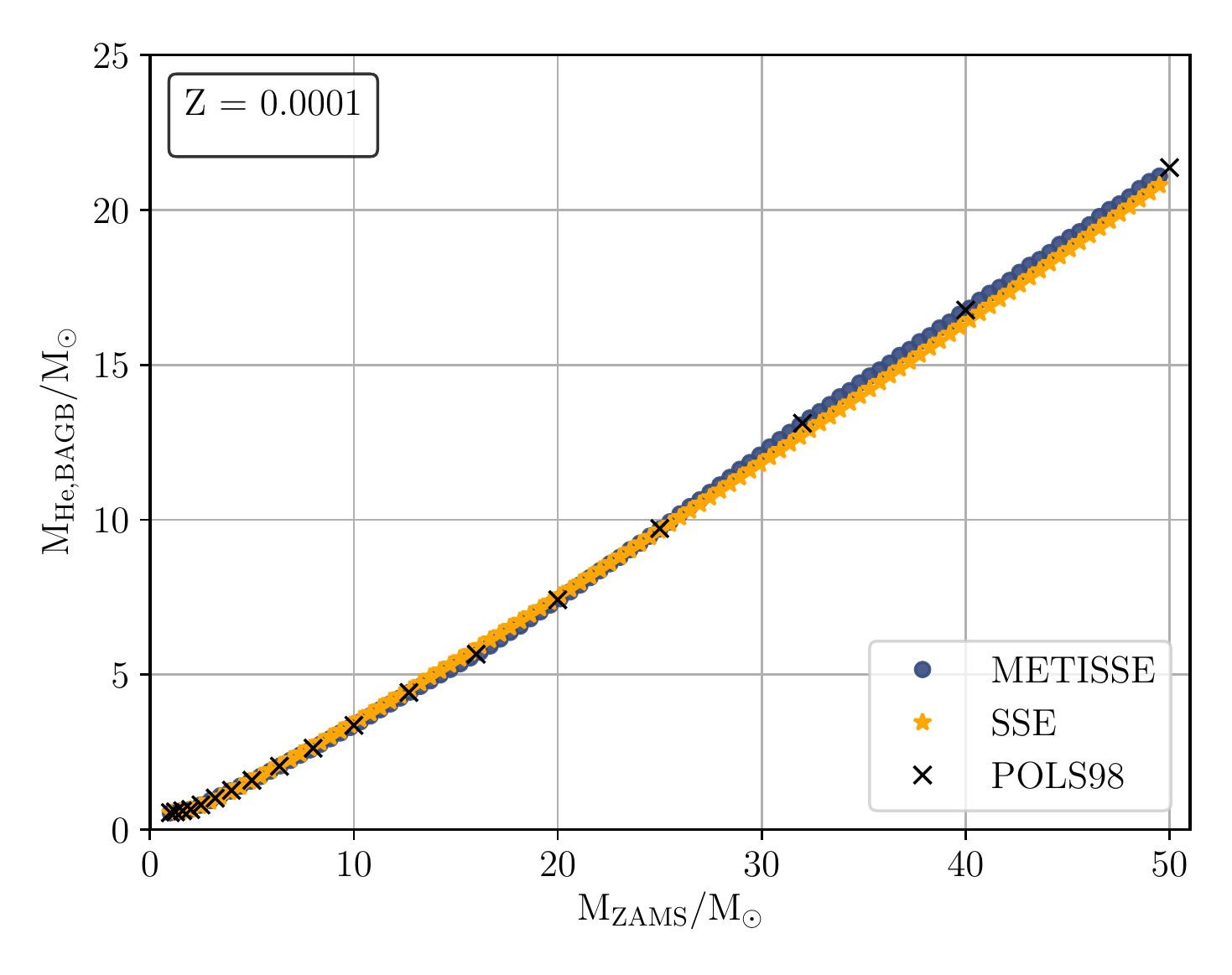}
    & 
    \includegraphics[width=0.9\columnwidth]{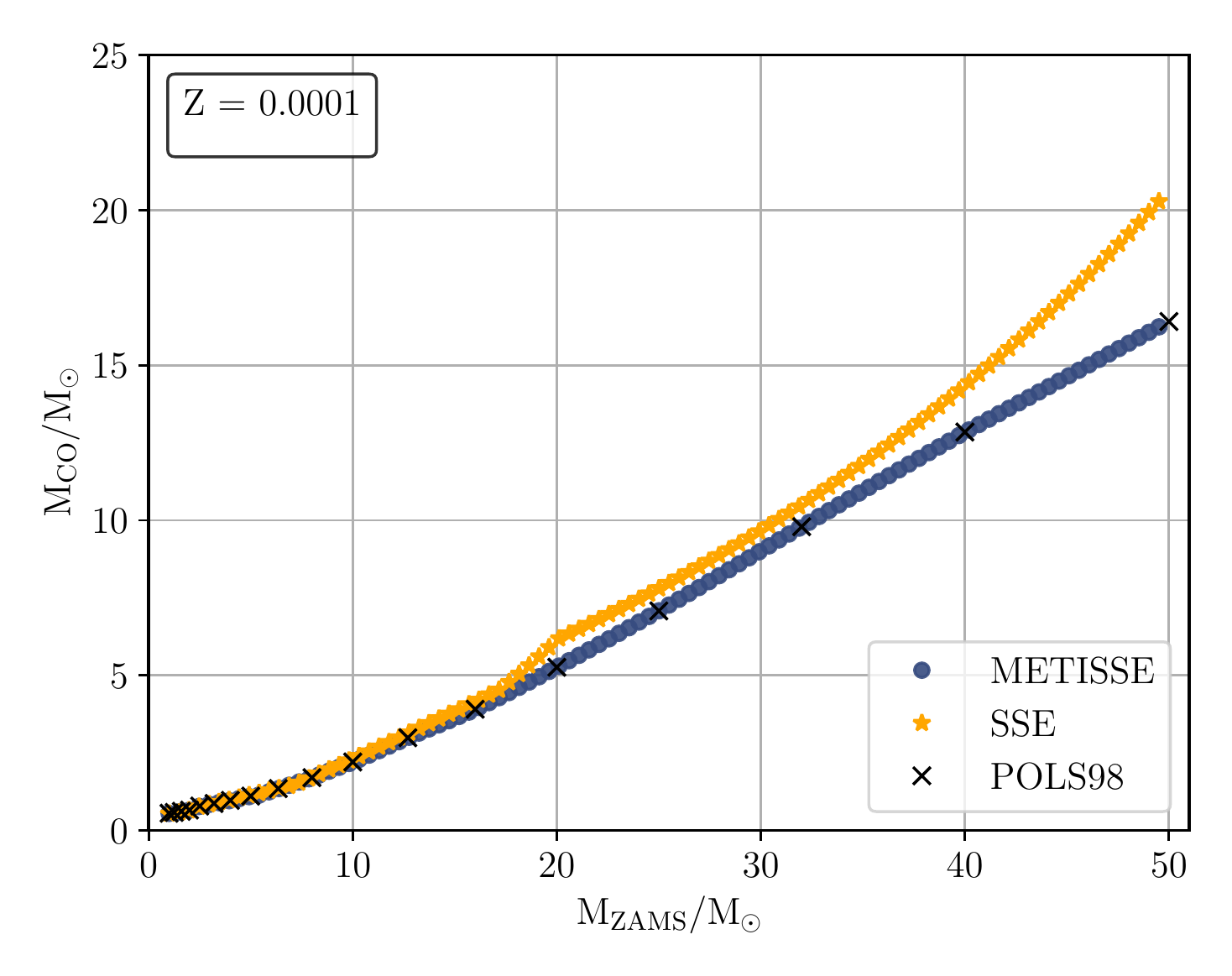}
    \\
    \end{tabular}
    \caption{He-core mass at the base of the AGB ($M_{\rm He,BAGB}$: left panels) and CO-core mass at the end of AGB ($M_{\rm CO}$: right panels) as a function of the ZAMS mass ($M_{\rm ZAMS}$) of the star for different metallicities (Z, as indicated in each panel).
    Star symbols show the values predicted by \textsc{SSE} while circles denote the values predicted by \textsc{METISSE} for a uniform distribution of stars of initial mass between 1 and 50\Msun{}, assuming no mass loss due to stellar winds. Corresponding values from \citet{1998MNRAS.298..525P} are marked as a cross.}
    \label{fig:remnant_SSE}
\end{figure*}

Any two methods of synthetic stellar evolution using the same input data should be able to produce matching output. Hence in Fig.~\ref{fig:SSE_METISSE}, we compare the tracks interpolated by \textsc{METISSE} using \citet{1998MNRAS.298..525P} models and tracks generated by \textsc{SSE} for the same input mass and metallicity ($Z = 0.02$). Because the set of stellar models used by the two codes is the same, the difference in the tracks simply reflects the difference between the use of fitting formulae and that of using interpolation. As is evident from the figure, \textsc{METISSE} is able to better preserve the finer details in the tracks, for example those during the Hertzsprung Gap. 

These seemingly tiny details in the tracks can lead to non-trivial dissimilarities in predicting other stellar properties. 
To show this, in Fig.~\ref{fig:remnant_SSE}, we plot the He core mass of stars at the base of the AGB (corresponding to the TAcHeB EEP) and the CO core mass at the end of AGB (corresponding to the TPAGB EEP for low and intermediate-mass stars, and C ignition for massive stars) as predicted by \textsc{METISSE} and by \textsc{SSE} for stars in the mass range 1 to 50\Msun{} with metallicities $Z = 0.02$ and $Z = 0.0001$. For $Z = 0.02$ the core masses predicted by \textsc{METISSE} agree well with \textsc{SSE}. There are some discrepancies in the prediction of CO core mass for stars with initial mass greater than about 40\Msun{}.
The differences are larger for lower metallicity ($Z = 0.0001$) and extend down to 20\Msun{} stars.

These differences are a result of how the evolution of the CO core is treated in each code. 
On the AGB, the CO core of a star grows in size owing to He-shell burning. If the star is massive enough, the core at some point can reach sufficient conditions to ignite carbon and the mass of the CO core can decrease. 
In \textsc{SSE}, the evolution of the CO core of a star has been simplified, allowing the CO core mass to grow until it reaches $M_{\mathrm{c}, \mathrm{SN}}$ (Equation~\ref{eqn:mcmax}). On the other hand, \textsc{METISSE} makes no prior assumptions and relies on the input set of detailed models to provide information about the CO core mass of the interpolated track. 
It can, therefore, more accurately relay the behaviour of the CO core that has been computed in the detailed input stellar models. This illustrates the reliability of stellar parameters computed with \textsc{METISSE}.

\subsection{Timing and performance}
\label{subsec:performance}

\begin{figure}
  \includegraphics[width=0.49\textwidth]{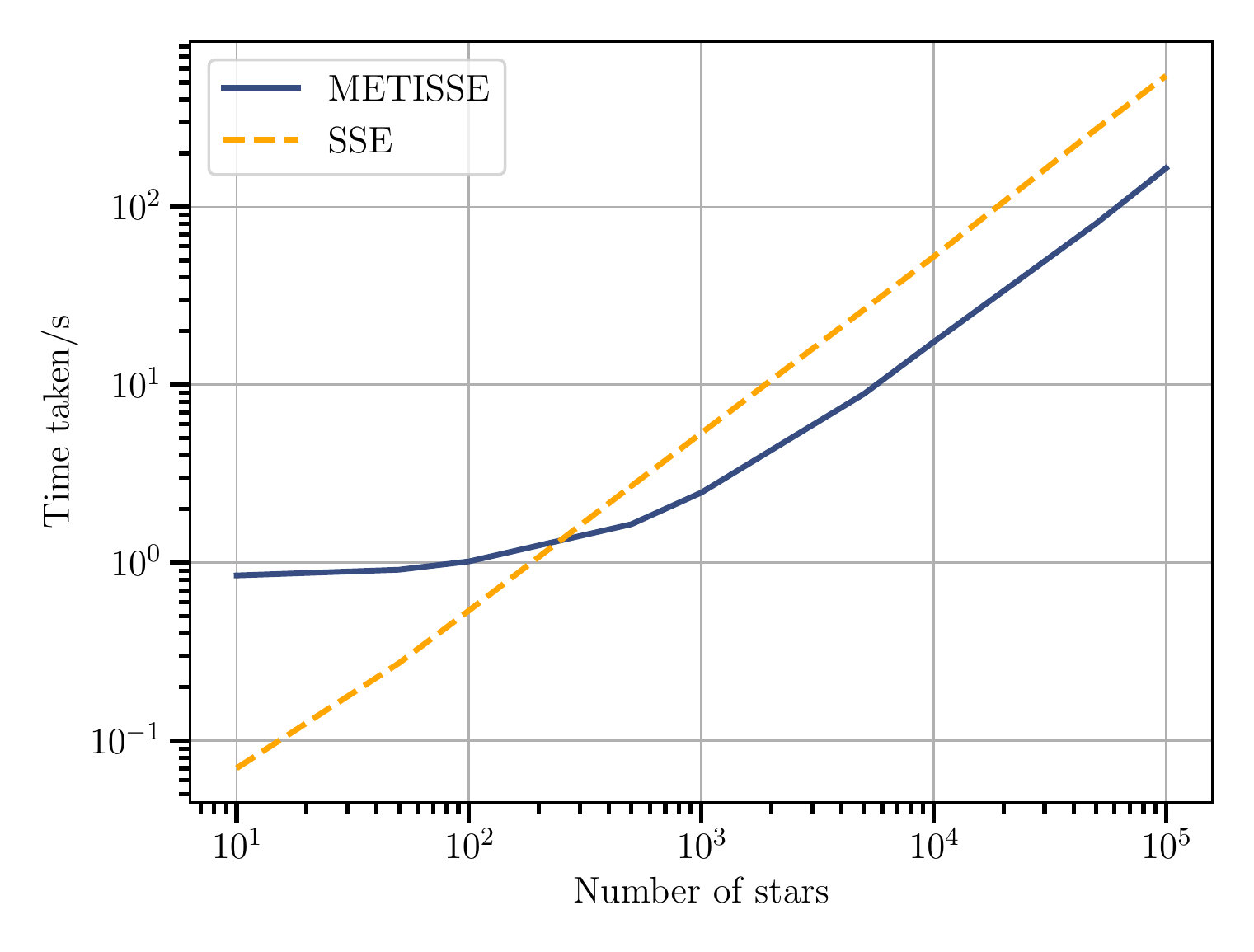}
  \caption{Timing \textsc{METISSE}: solid line represents the time taken by \textsc{METISSE} while the dashed line is the time taken by \textsc{SSE} as a function of number of stars evolved. The timing is for a single 2.3 GHz Intel i5 core.}
  \label{fig:time_METISSE}
\end{figure}

In \textsc{METISSE}, input tracks from the chosen detailed evolution code need to be read and loaded in the computer memory before any interpolation can be performed. Depending on the density of the input set of models, the memory requirement can be of the order of Megabytes to Gigabytes. 
The memory required depends not only on the number of tracks but also on the amount of data read for each track.

The number of data columns from the input tracks can be easily controlled by the user in \textsc{METISSE}. By selecting fewer columns, one can speed up the runs and reduce memory usage. This is useful for simulating systems with millions of stars (e.g. globular clusters in N-body simulations). If more surface abundances are needed, for example, to trace the evolution of different elements in stellar populations, the columns can be included from the detailed stellar models with only a modest increase in the memory usage and computing time.

To compare the performance of \textsc{METISSE} with \textsc{SSE}, we computed 10 to 10$^{5}$ stellar tracks between 1 and 50\Msun{}, evolving each star up to $12\,$Gyr for each method. For a fair comparison, the input set of tracks and data columns used by \textsc{METISSE} were kept the same as in \textsc{SSE}. 
In Fig.~\ref{fig:time_METISSE}, we show the average time taken by \textsc{METISSE} compared to that by \textsc{SSE} to evolve different numbers of stars. For \textsc{SSE} the increase in run-time with the number of stars is linear. \textsc{METISSE} requires more time (0.8\,s here) in the beginning to process the set of input tracks, independent of the number of stars evolved. Hence, for fewer stars, \textsc{METISSE} takes longer than \textsc{SSE} to complete the run. For larger populations however, the time taken to process the input tracks becomes a negligible fraction of the total run time and \textsc{METISSE} becomes almost three times faster than \textsc{SSE}.

It is necessary to emphasize here that like memory, the time taken by \textsc{METISSE} does increase depending on the number of input stellar tracks. Overall, it can be safely concluded that at the very least \textsc{METISSE} is comparable to \textsc{SSE} in terms of performance.

\section{\textsc{METISSE} with \textsc{MESA} and \textsc{BEC}: Predicting properties of massive stars}
\label{sec:compare_stellar_models}

\begin{figure*}
\begin{tabular}{cc}
\includegraphics[width=0.9\columnwidth]{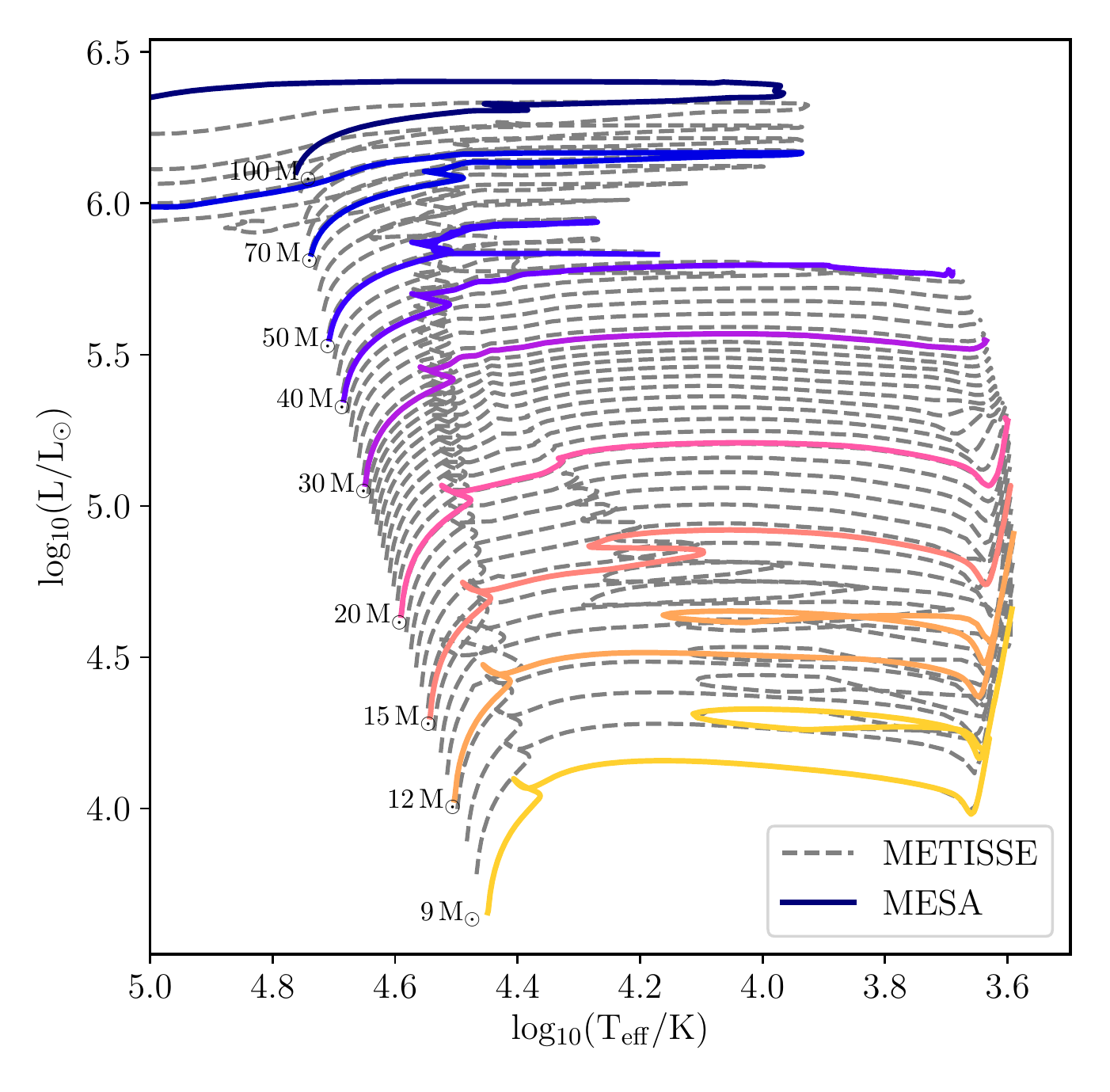}
&
\includegraphics[width=0.9\columnwidth]{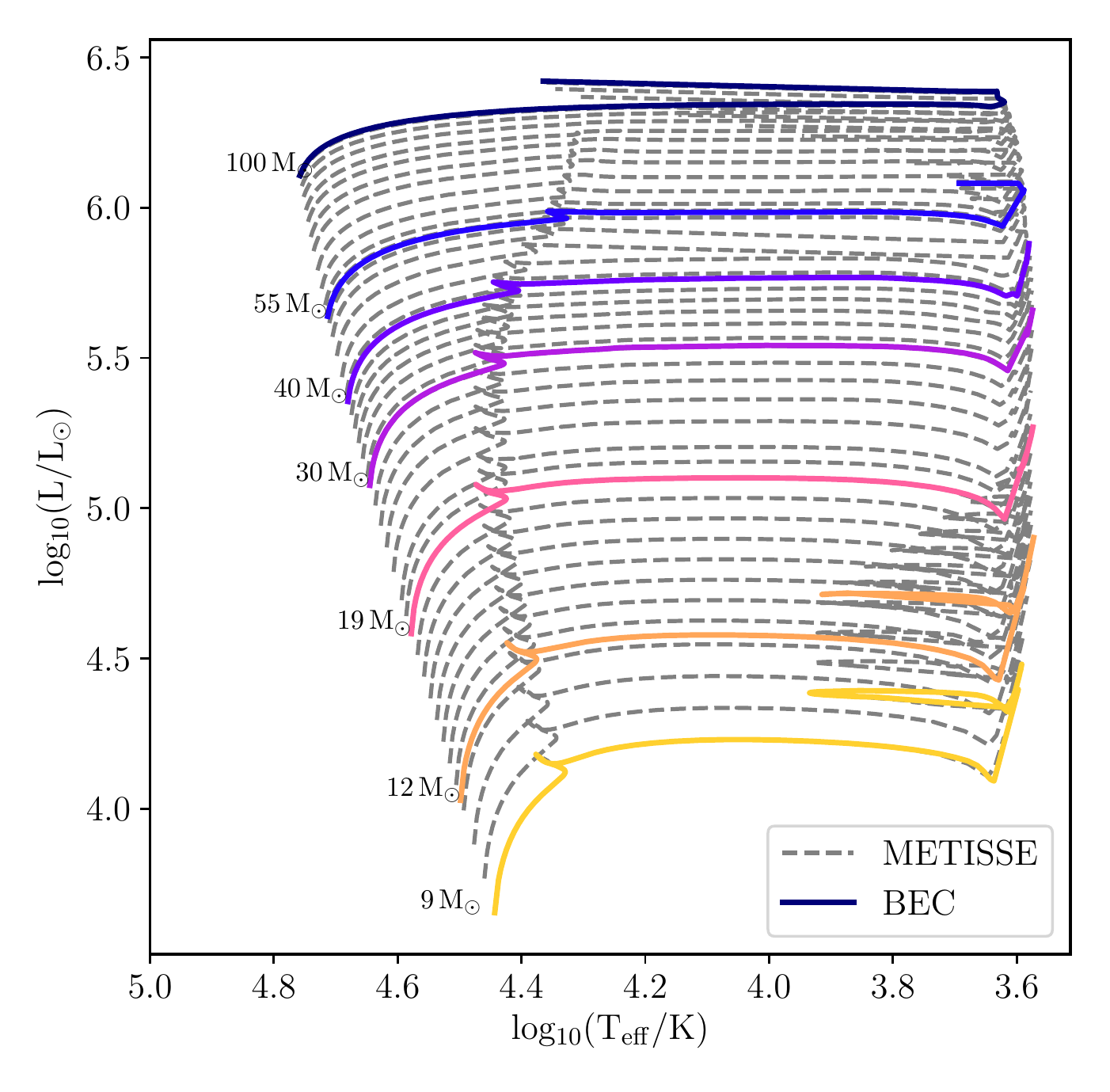}
\\
\end{tabular}
\caption{HR diagram showing tracks in the mass range 9 to 100\Msun{} interpolated with \textsc{METISSE} with detailed tracks from \textsc{MESA} (for $Z = 0.00142$, left panel) and tracks from \textsc{BEC} (for $Z = 0.00105$, right panel) as input. For clarity, only selected tracks from each set are shown.}
\label{fig:metisse_interp}
\end{figure*}

Massive stars are responsible for the chemical enrichment of their surroundings. They are precursors of astrophysical transient phenomena including supernovae and gamma-ray bursts, progenitors of compact objects.
As these stars are rare in nature, their evolutionary parameters,
such as mass-loss rates, mixing processes and nuclear reaction rates, are not very well constrained \citep[see, for example,][]{2016ApJS..227...22F, 2017A&A...603A.118R, 2018ApJS..234...19F}. Therefore stellar evolution codes make certain assumptions about the interior and physics of these stars which can lead to different evolutionary outcomes. In order to check the validity of these assumptions, it is necessary to compare their predictions with observations of massive stellar populations. 
For this one needs to be able to apply different stellar evolution models in population synthesis codes. 

Built exactly for this purpose, \textsc{METISSE} can read different sets of evolutionary tracks, including those generated by different stellar evolution codes. The only requirement is that the input tracks should be in the EEP format. 
In this section, we demonstrate the capability of \textsc{METISSE} to use sets of evolutionary tracks evolved using \textsc{BEC} and \textsc{MESA}. 
We apply the sets of stellar models introduced in Sections~\ref{subsec:MESA_stellar_models} and \ref{subsec:BEC_stellar_models} respectively, as an input to \textsc{METISSE} and interpolate 100 stars uniformly distributed in mass between 9 and 100\Msun{} at metallicity $Z = 0.00142$ for \textsc{MESA} tracks and $Z = 0.00105$ for \textsc{BEC} tracks. The HR diagram for a subset of both the detailed and interpolated tracks is shown in Fig.~\ref{fig:metisse_interp}. We use the results presented in this section to explore the impact of stellar evolution parameters on the evolution of massive stars.

We also compare said outcomes to those obtained using \textsc{SSE} for $Z = 0.00142$. For \textsc{SSE} the maximum mass of the detailed tracks used for calculating the fitting formulae was about 50\Msun{}. Stars above this mass are calculated by extrapolating the fitting formulae from less massive stars. Moreover, detailed tracks of \citet{1998MNRAS.298..525P} do not include wind mass loss. Consequently, mass-loss in \textsc{SSE} is modelled by removing the mass from the stellar envelope. We have used the mass-loss rates of \citet{2010ApJ...714.1217B} in the \textsc{SSE} tracks presented here.

\subsection{Impact on remnant mass}
\label{subsec:impact_on_remnnant_mass}

Massive stars are the progenitors of compact objects: neutron stars and black holes whose mergers result in the emission of gravitational waves observable by LIGO/Virgo \citep{2016PhRvL.116f1102A}. Therefore, the ability to accurately predict stellar remnant masses is crucial. 
The remnant masses can be calculated from the total mass and the core properties of the stars using prescriptions such as those in \citet{2012ApJ...749...91F}.

For tracks interpolated with \textsc{METISSE} using \textsc{MESA} and \textsc{BEC} models, we calculate the mass of stellar remnants in the manner outlined in Section~\ref{subsec:stellar_remnants}. We have followed \citet[][]{2008ApJS..174..223B} for calculating the mass of remnants \citep[same as StarTrack prescription in][]{2012ApJ...749...91F}. For stars with final CO core mass less than 5\Msun{}, the prescription yields a remnant mass based on the iron-nickel (FeNi) core mass of the star while for stars with CO cores more massive than 7.6\Msun{}, it is assumed that the whole star collapses to form a black hole. In between the two regimes, partial fallback from the star is assumed and the mass of the remnant follows a linear fit between the FeNi core mass and the total mass of the star. The FeNi core mass itself is calculated from the CO core mass. To account for mass lost due to neutrino cooling of stellar cores before the supernova explosion, the baryonic mass of the remnant obtained above is converted to its gravitational mass \citep[following Equations~3 and~4 of][]{2008ApJS..174..223B}. This simple model assumes no mass gap \citep[][]{2010ApJ...725.1918O, 2011ApJ...741..103F} between neutron stars and black holes. 

Following \citet{2008ApJS..174..223B} we suppose the maximum neutron star mass to be 3\,M\sun{} in this work, although the maximum observed is 2.14\Msun{} \citep{2020NatAs...4...72C}. 
The relationship between core mass and remnant mass may not follow this simple relation; recent works have suggested that in some mass ranges, certain stars may form neutron stars while others form black holes \citep[e.g.][]{Sukhbold:2013yca, Ertl:2015rga, 2020MNRAS.492.2578S}.

\begin{figure}
  \includegraphics[width=\columnwidth]{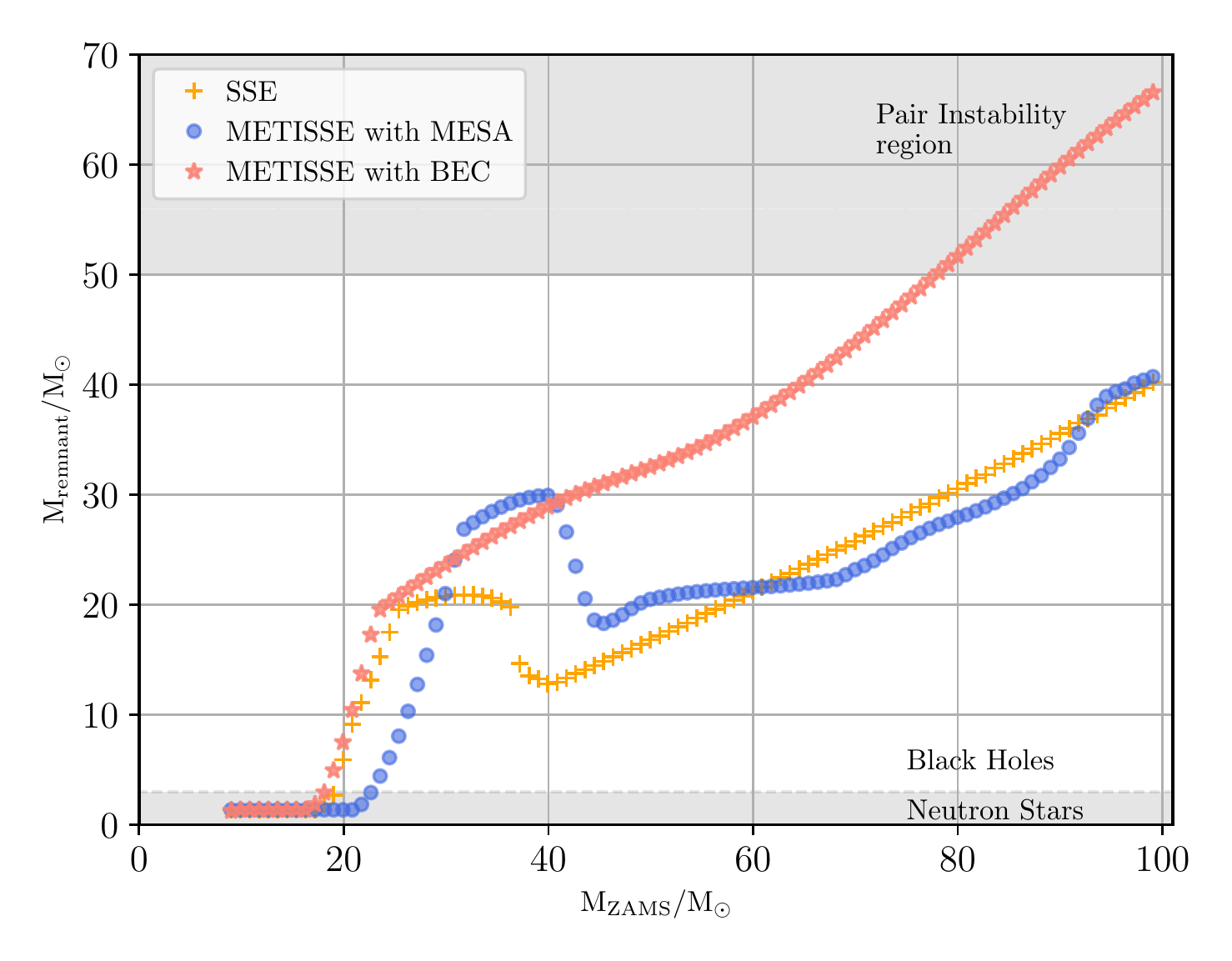}
  \caption{The mass of stellar remnants versus the mass of their progenitors, as predicted by \textsc{SSE} (yellow crosses), \textsc{METISSE} with \textsc{MESA} (blue circles) and \textsc{METISSE} with \textsc{BEC} (red stars). The grey area above 50\,M\sun{} shows the region where stars may encounter pair instability. See section~\ref{subsec:impact_on_remnnant_mass} for details.}
  \label{fig:rem}
\end{figure}

In Fig.~\ref{fig:rem}, we plot the results in terms of remnant mass obtained using \textsc{SSE}, \textsc{METISSE} with \textsc{MESA} models and \textsc{METISSE} with \textsc{BEC} models against ZAMS mass of their progenitors. 
For the \textsc{BEC} models, stars with initial masses greater than 80\Msun{} have final core masses greater than 50\Msun{}. Stars with core masses from about 50 to 130\Msun{} are expected to encounter the well-known pair instability condition during their post-He-burning evolution (typically during O-burning), leading to enhanced mass loss or total destruction of the star \citep{FowlerHoyle:1964ApJS,1968Ap&SS...2...96F,Woosley:2016hmi, 2019ApJ...882..121S}. Currently, we do not take into account the effect of pair instability or pulsational pair instability when predicting remnant masses in \textsc{METISSE} but, for reference, the region where pair instability becomes relevant is highlighted in the figure. 

We find that there is a striking variation in remnant mass predicted by \textsc{SSE} and both the \textsc{MESA} and \textsc{BEC} models in \textsc{METISSE}. For stars with ZAMS mass between 9 and 18\,M\sun{}, the three sets of tracks agree well. For stars with ZAMS masses between 19 and 30\Msun{}, there is a linear increase in the remnant mass owing to partial fall-back of matter on to the collapsing core during the supernova explosion. For \textsc{MESA} tracks the rise in remnant mass is slower than the other two sets of tracks and peaks at around a 40\Msun{} ZAMS mass while for \textsc{SSE} the local maximum occurs around 30\Msun{}. \textsc{BEC} tracks do not show any such decline and the difference in the remnant mass between \textsc{MESA} and \textsc{BEC} becomes pronounced (about 20\Msun{}) for stars with a ZAMS mass more than 40\Msun.

The mass of the remnant is clearly influenced by the choice of stellar models and the different choices of stellar parameters adopted therein. We discuss these differences, their origins and their impact on the remnant masses in Section~\ref{sec:discussion_of_results}. 

\subsection{Impact on radius evolution}
\label{subsec:impact_on_radius_evolution}
\begin{figure}
  \includegraphics[width=\columnwidth]{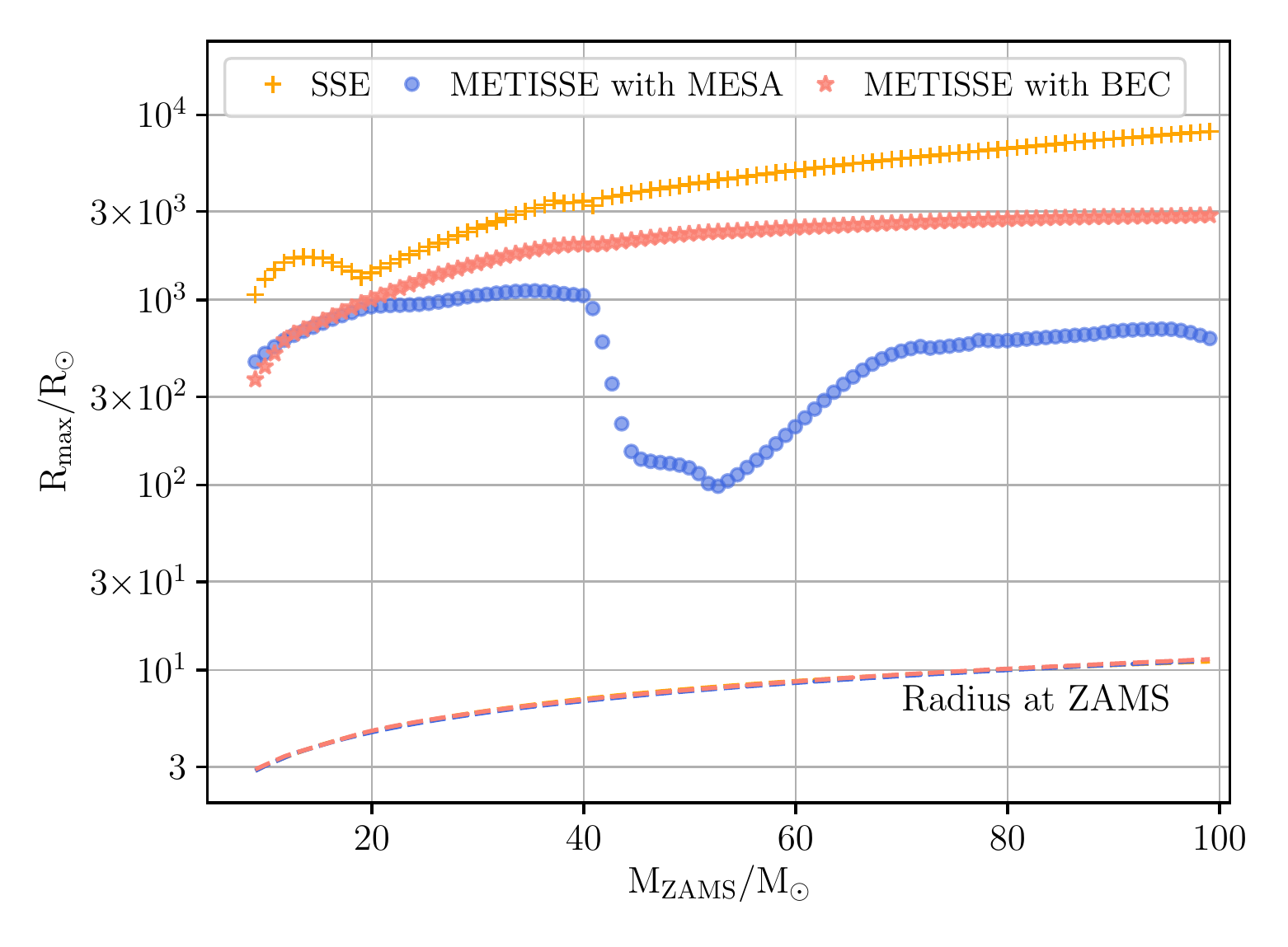}
  \caption{Maximum radii obtained by stars as a function of their ZAMS mass, symbols are the same as in Fig.~\ref{fig:rem}. The dashed lines (indistinguishable here) represent the ZAMS radius for each of the three sets.}
  \label{fig:max_rad}
\end{figure}

Most stars expand as they evolve, becoming giants. This is especially important for stellar evolution in binary systems because the expanding star can fill its Roche lobe and initiate a phase of mass transfer in the system. Hence, accurately predicting the extent of radial expansion for a star is necessary to determine the evolution of binary systems.

In Fig.~\ref{fig:max_rad}, we plot the maximum radii of stars uniformly distributed in mass between 9 and 100\Msun{} calculated with \textsc{MESA} and \textsc{BEC} models in \textsc{METISSE} and with \textsc{SSE}. Similar to Fig.~\ref{fig:rem}, there is disparity between the results obtained with the three sets of tracks. For \textsc{SSE} and \textsc{BEC} the maximum radial expansion achieved by the stars increases with initial mass (aside from a slight decrease for \textsc{SSE} near 15\Msun{}). For \textsc{MESA} tracks however, the trend changes considerably beyond 40\Msun{}: the maximum radius decreases until 55\Msun{} reaching a minimum of about 100\,R\sun{} before slowly increasing for more massive stars.

The lower radii predicted by the various models impact the outcome of close binary interaction. The number of interacting binaries with orbital separations that lie within the range between the minimum radius (${\rm R_{min}}$) and the maximum radius (${\rm R_{max}}$) of the star can be given by
\begin{equation}
    N=\int_{R_{\min}}^{R_{\max}} \frac{dN}{da}da \, .
    \label{eq:N1}
\end{equation}

Assuming a distribution of binary orbital separations $a$ that is flat in $\log a$  \citep{1924PTarO..25f...1O,1983ARA&A..21..343A}, $dN/da \propto 1/a$, for Equation~\ref{eq:N1} we can write
\begin{equation}
\begin{split}
    N   &\propto \int_{R_{\min}}^{R_{\max}} \frac{1}{a}da 
        &= [\ln a]_{R_{\min} }^{R_{\max}} 
        &= \ln R_{\max}-\ln R_{\min} \, .
\end{split}
\end{equation}

Therefore, the ratio between the number of interacting binaries predicted by for example \textsc{MESA} to \textsc{SSE} can be given as
\begin{equation}
    \frac{N^{\rm {SSE}}}{N^{\rm {MESA}}} = \frac{\ln R_{\max} ^{\rm SSE}-\ln R_{\min} ^{\rm SSE}}{\ln R_{\max}^{\rm MESA}-\ln R_{\min}^{\rm MESA}}  \, .
    \label{eqn:ratio}
\end{equation}

We applied Equation~\ref{eqn:ratio} to each stellar track in the three sets. On average, \textsc{SSE} predicts 1.6 times more interacting binaries than \textsc{METISSE} with \textsc{MESA}. Doing a similar exercise using the \textsc{BEC} tracks, we find that \textsc{SSE} predicts 1.3 times more interacting binaries than \textsc{METISSE} with \textsc{BEC}. Both numbers are comparable to differences in uncertainties in the initial conditions of binaries \citep[][]{2015ApJ...814...58D,2018A&A...619A..77K}. 
The difference can be larger for the most massive stars. E.g. for a 60\Msun{} star, \textsc{SSE} predicts 2.3 times more interacting binaries than \textsc{MESA}, and 1.4 times more than \textsc{BEC}.
However, to account for the fact that massive stars are less common in nature, we weight the above average by an initial mass function with a power law index of $\alpha = -2.3$ for masses above 1\Msun{} \citep{Salpteter:1955ApJ,Kroupa:2000iv}. For this more realistic population of binaries, \textsc{SSE} still predicts 1.25 times more interacting binaries than \textsc{MESA} and 1.18 times more interacting binaries than \textsc{BEC}. We further discuss the origin of these differences Section~\ref{sec:discussion_of_results}.

\subsection{Impact on main-sequence lifetime}

\begin{figure}
  \includegraphics[width=\columnwidth]{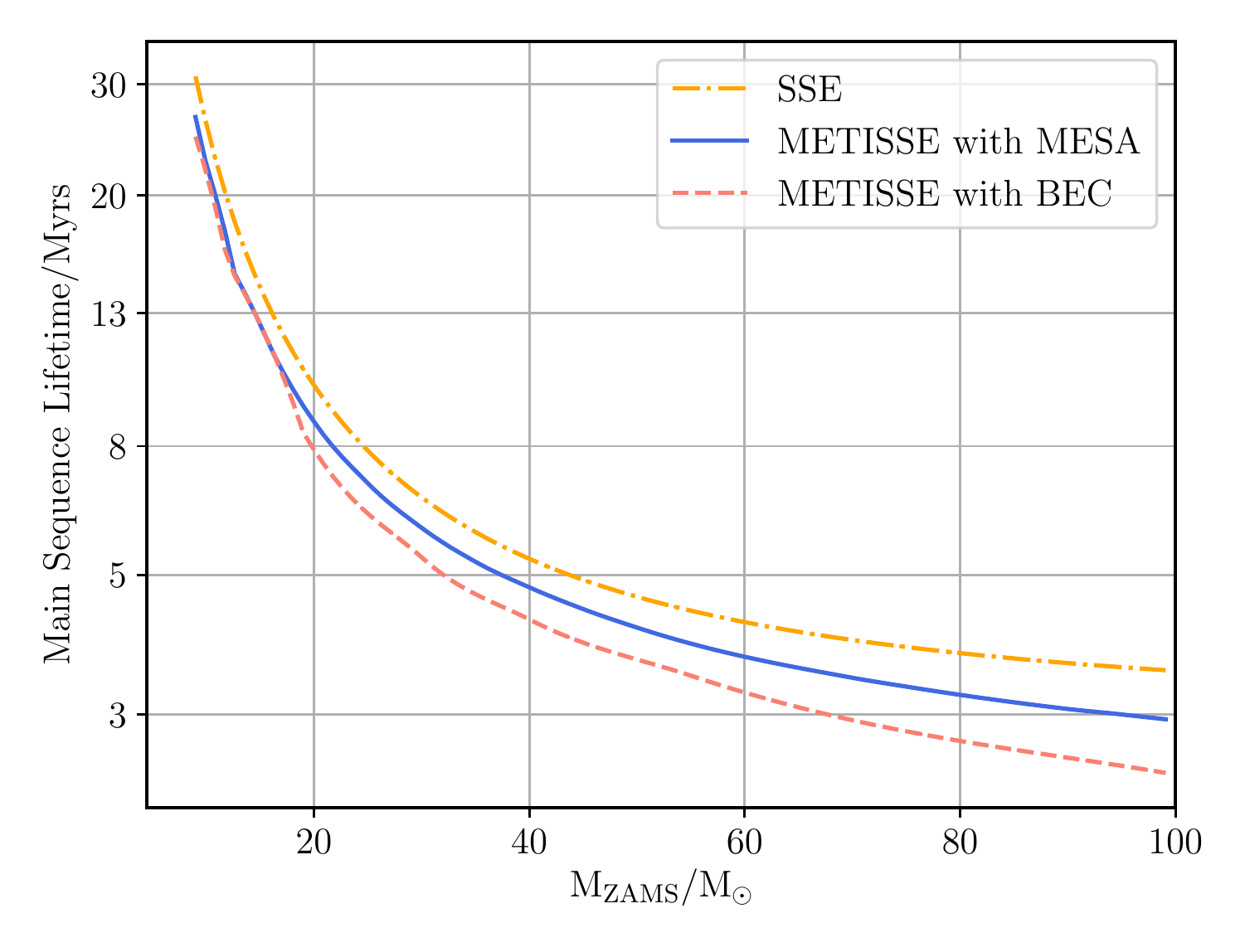}
  \caption{Main-sequence lifetime of stars as a function of their ZAMS mass, as predicted by \textsc{SSE} (dashed dotted line), \textsc{METISSE} with \textsc{MESA} tracks (solid line) and \textsc{METISSE} with \textsc{BEC} (dashed line).}
  \label{fig:ms_lifetime}
\end{figure}

Young massive star clusters are instrumental in the study of stellar dynamics and the stellar mass function \citep*{2010ARA&A..48..431P}. A key method for determining the age of star clusters is to use the main-sequence turnoff age which requires estimation of the main-sequence lifetime of stars \citep{1998MNRAS.298..525P,2010RSPTA.368..755K}.  
The MS lifetime can differ between models owing to the difference in the treatment of mixing processes inside the star.
Processes like convection and overshooting can help replenish H supply in the core, prolonging the time spent in the MS phase. Mixing parameters are often calibrated using values from a solar model and might not be applicable to massive stars \citep{2018ApJ...856...10J}. Differences in the MS lifetimes of massive stars, as predicted by different sets of tracks, can be useful to explain phenomena such as extended main-sequence turnoffs and the age spread observed in young massive star clusters \citep{2001MNRAS.324..367J, 2017ApJ...844..119L}.

In Fig.~\ref{fig:ms_lifetime}, we plot the time spent on the MS by stars of mass 9 to 100\Msun{} as predicted by \textsc{SSE}, \textsc{METISSE} with \textsc{MESA} and \textsc{METISSE} with \textsc{BEC}. The difference in predicted lifetimes varies from about 0.5~Myr for a 40\Msun{} star to about 4~Myr for a 9\Msun{}, between each set. This corresponds to roughly 10 to 20 per cent of the total time spent in the MS phase. 
In Section~\ref{subsec:disc_conv}, we discuss the effect on the MS lifetimes arising from differences between the treatment of convection and the choice of the overshooting parameters adopted in the input stellar models.

\section{Understanding the differences between input stellar models}
\label{sec:discussion_of_results}

As \textsc{METISSE} relies on having an input set of detailed models to provide information about the interpolated track, the difference in the properties of massive stars obtained with \textsc{MESA} and \textsc{BEC} models in \textsc{METISSE} pointed out in Section~\ref{sec:compare_stellar_models} can be attributed to the input parameters employed while computing the detailed stellar models. 

In this section we discuss the role of three major contributors (i) modelling of radiation dominated envelopes of massive stars, (ii) mass-loss rates and (iii) convection and overshooting parameters. Although other factors such as rotation, chemical composition and surface boundary conditions can also have an impact on the structure and evolution of massive stars, the discussion of these requires dedicated future studies. 

\subsection{Massive stellar envelopes and the role of the Eddington luminosity}
\label{subsec:disc_edd_lum}
    
\begin{figure*}
    \centering
    \begin{minipage}{0.49\textwidth}
    \centering
    \includegraphics[width=\textwidth]{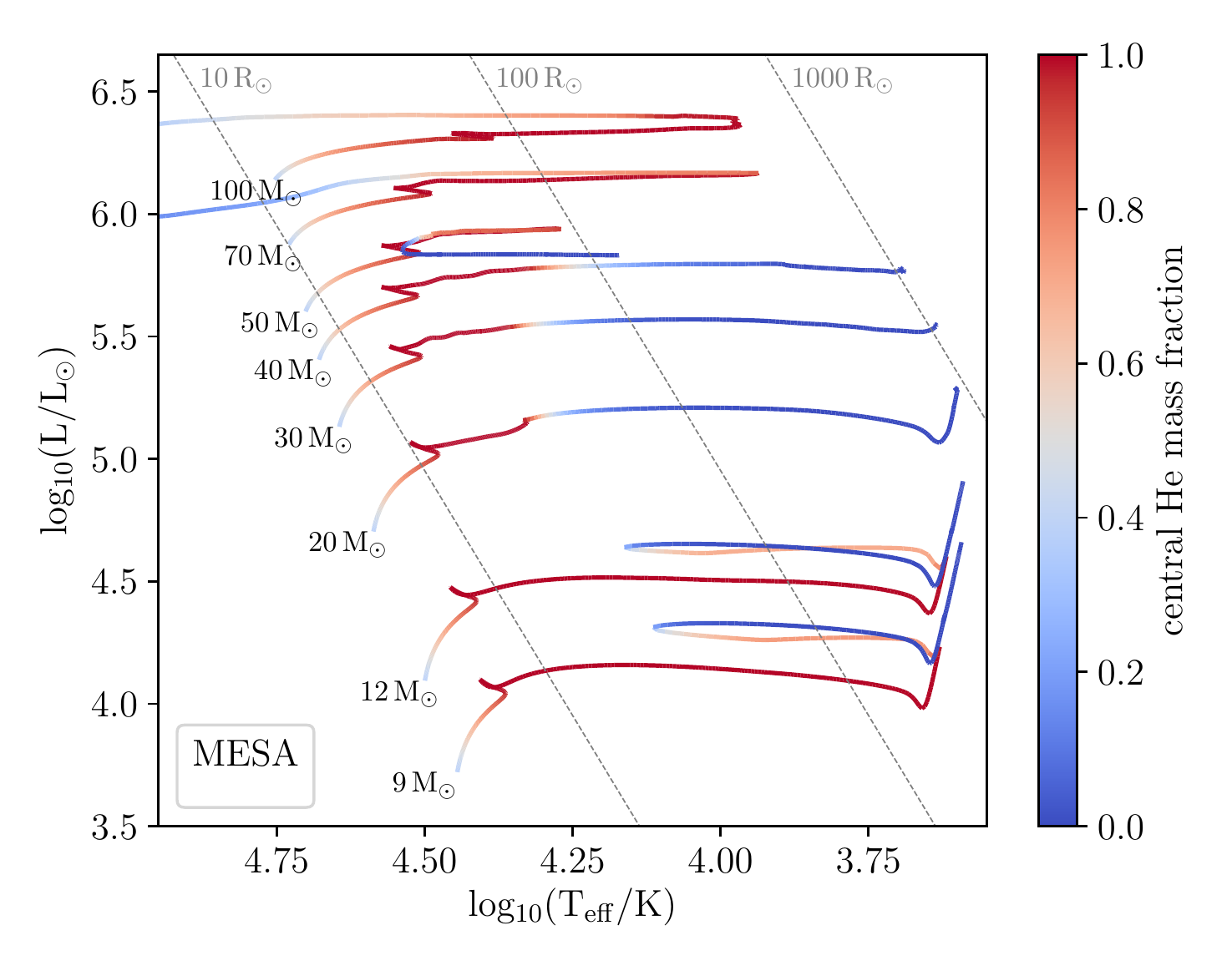}
    \end{minipage}
      \hfill
    \begin{minipage}{0.49\textwidth}
    \centering
    \includegraphics[width=\textwidth]{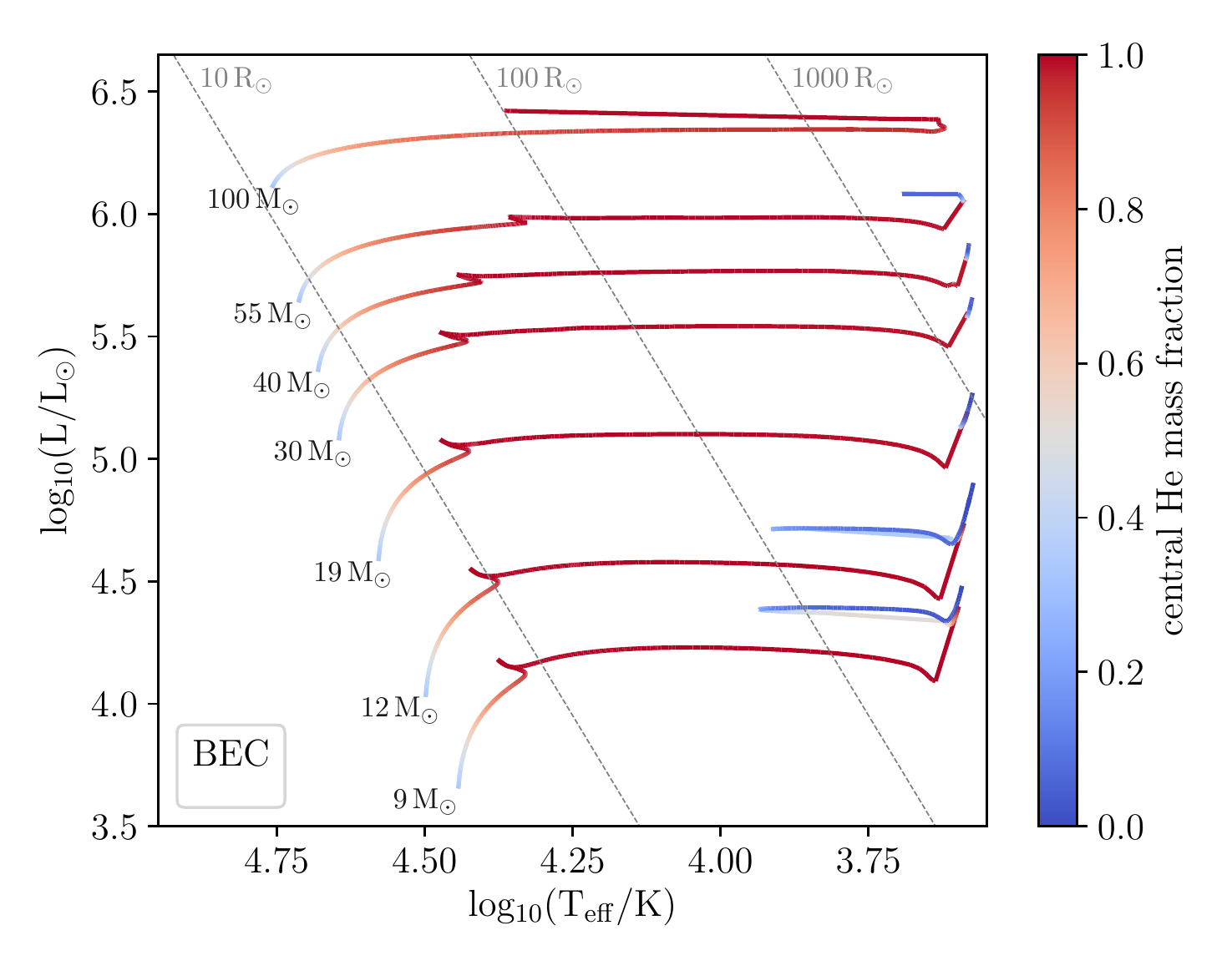}
    \end{minipage}
    \caption{HR diagrams showing stellar tracks evolved using \textsc{MESA} (left) and \textsc{BEC} (right) and coloured according to their central helium mass fraction. For clarity, only nine (out of the 25 computed in this work) \textsc{MESA} tracks are shown here. The differences in the tracks are due to different physical inputs, as discussed in Section~\ref{sec:discussion_of_results}.}
  \label{fig:He_fraction}
%\vspace{-0.5in}
\end{figure*}

The Eddington limit for a spherically symmetric star in hydrostatic equilibrium is defined as the maximum outward radiative motion of stellar material that can be balanced by the inwards acting gravitational force \citep*{1926ics..book.....E,2004ApJ...616..525O}. For a star containing mass $m(r)$ inside radius $r$ and radiative opacity $\kappa(r)$, the expression for the Eddington luminosity is given by
\begin{equation}
    L_{\rm {Edd}}(r)=\frac{4 \pi c G m(r)}{\kappa(r)} \, .
    \label{eq:ledd}
\end{equation}

Hence a critical limit, known as the Eddington factor can be defined \citep[following][]{1997ASPC..120...83L} as 
\begin{equation}
    \Gamma=\frac{L(r)}{L_{\rm {Edd}}(r)} = \frac{\kappa(r)}{4 \pi c G} \frac{L(r)}{m(r)} \, .
    \label{eq:gamma}
\end{equation}

For massive stars, the luminosity inside the stellar envelope can exceed the Eddington limit $(\Gamma >1)$ due to elemental opacity peaks \citep*{1992ApJ...397..717I,2009A&A...499..279C}, e.g, towards the end of the main sequence. Moreover, the outer envelopes of massive stars are dominated by radiation pressure and the convective transport of energy given by standard Mixing Length Theory (MLT; see Section~\ref{subsec:disc_conv}) becomes inefficient. As shown by \citet*{1973ApJ...181..429J}, $\Gamma >1$ combined with inefficient convection can lead to pressure and density inversion inside the stars, $dp/dr >0$ and $d\rho/dr >0$. This means that density and gas pressure increases outwards for massive stars with super-Eddington luminosity in their outer envelopes and can cause numerical difficulty in modelling stars with 1D stellar evolution codes \citep{2013ApJS..208....4P}. To push the evolution of a star beyond this point, 1D stellar evolution codes adopt different approximations. 

In the \textsc{BEC} stellar models, density and pressure inversion inside the stellar envelope causes the hydrostatic expansion of the outermost layers of the star \citep[envelope inflation,][]{2015A&A...580A..20S, 2017A&A...597A..71S}. The stellar models develop an extended, tenuous envelope in response to temperature and density inversions until the Eddington limit is no longer exceeded. The star becomes a supergiant\footnote{As pointed out by \citet{Szecsi:2015}, core-hydrogen burning cool supergiants are different from the usual red supergiants which expand in response to H-shell burning.} even while burning hydrogen in the core which affects its structure and evolution. 
The small time-steps required to resolve the inflated envelope of a star on the hydrodynamical time-scale pose a numerical difficulty for the post-main-sequence evolution \citep{2015A&A...580A..20S}. The \textsc{BEC} track with initial mass of 100\Msun{} here has been post-processed in the framework of the BoOST project to include a smooth approximation of the core helium burning phase \citep[see][for details]{szcsi2020bonn}.

In \textsc{MESA}, the density and pressure inversion can be mitigated through a formalism known as MLT++. For each model, \textsc{MESA} calculates \citep[cf. equation 38 of][]{2013ApJS..208....4P}:
\begin{equation}
\lambda_{\max } \equiv \max \left(\frac{L_{\mathrm{rad}}}{L_{\mathrm{Edd}}}\right) \quad \text { and } \quad \beta_{\min } \equiv \min \left(\frac{P_{\mathrm{gas}}}{P}\right) \, .
\label{eq:mlt++}
\end{equation}

Based on these parameters, \textsc{MESA} can artificially decrease the superadiabaticity (the difference between the isothermal and adiabatic temperature gradients) when stars approach their Eddington limits. Adopting the MLT++ formalism helps with the convergence of the models. However it can modify the radius and luminosity of the star and hence affect the mass-loss rates \citep{2013ApJS..208....4P}. \textsc{MESA} models in this work make use of the MLT++ formalism. Radiative pressure at the surface of the star is also enhanced to help with convergence of the model.

To investigate the effect of Eddington limit proximity on the stellar models, we plot the detailed stellar tracks from \textsc{MESA} and \textsc{BEC} in Fig.~\ref{fig:He_fraction}. Each track is coloured based on the He fraction in the centre of the star to show the location of the star during core He burning. The figure shows that stars evolved with \textsc{MESA} burn helium at higher temperatures and smaller radii compared to \textsc{BEC} where stars burn He at lower temperatures and larger radii.

With \textsc{MESA}, a 50\Msun{} star approaches the Eddington limit at the end of the main sequence. The proximity to the Eddington limit causes the star to experience high mass-loss rates which expose the hotter inner layers and the star moves bluewards in the HR diagram. The onset of H-shell burning causes the star to expand, lowering the surface temperature, making it lose even more mass. Therefore, local minima for both remnant masses and maximum radii are encountered for the tracks interpolated from \textsc{MESA} models in this region (i.e. about 45 to 55\Msun{} stars in Fig.~\ref{fig:rem} and Fig.~\ref{fig:max_rad}). Stars more massive than 60\Msun{} lose their envelope in \textsc{MESA} and become naked He star before they can finish burning He. 

\begin{figure*}
    \centering
    \begin{tabular}{cc}
     \includegraphics[width=0.9\columnwidth]{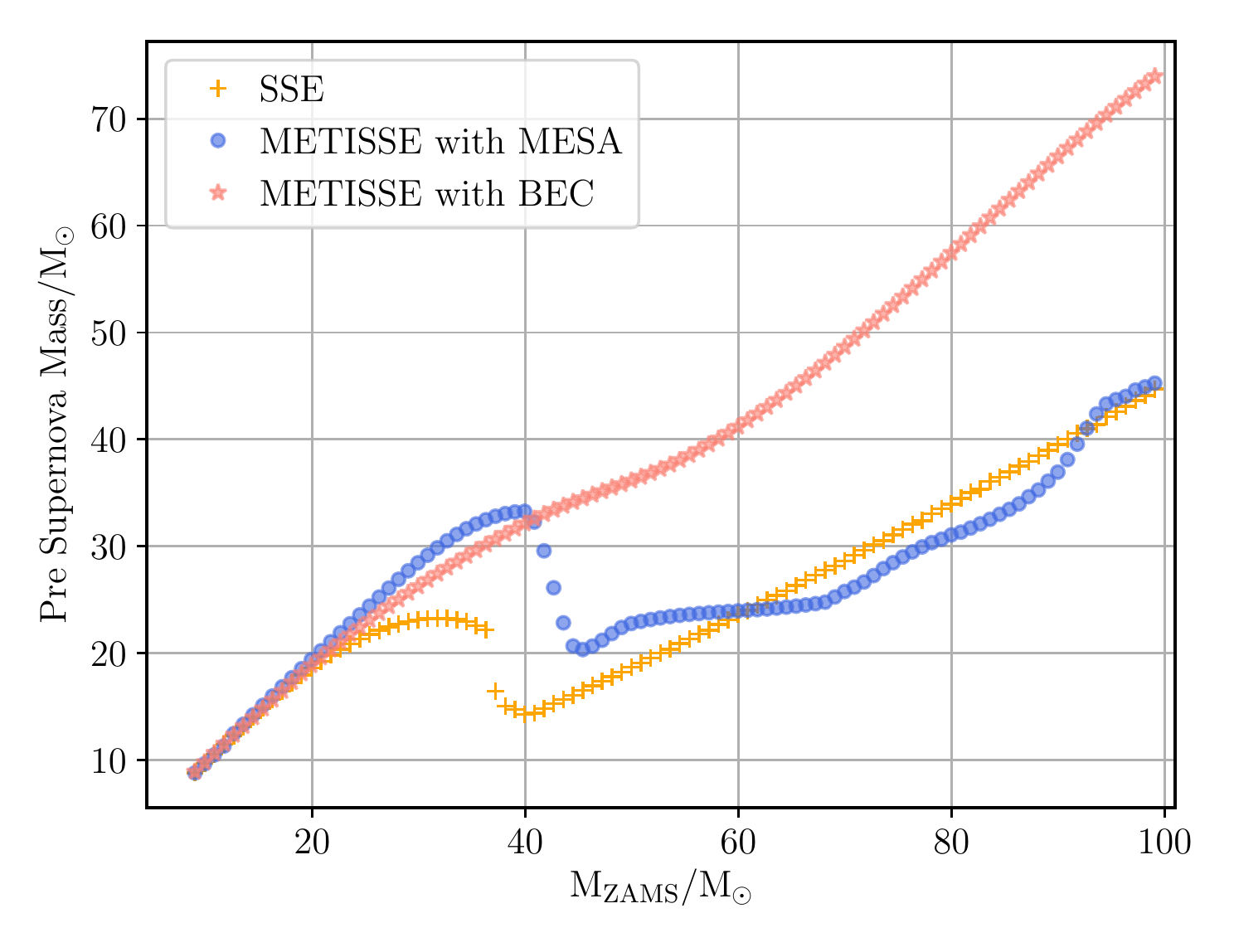}
    &
    \includegraphics[width=0.9\columnwidth]{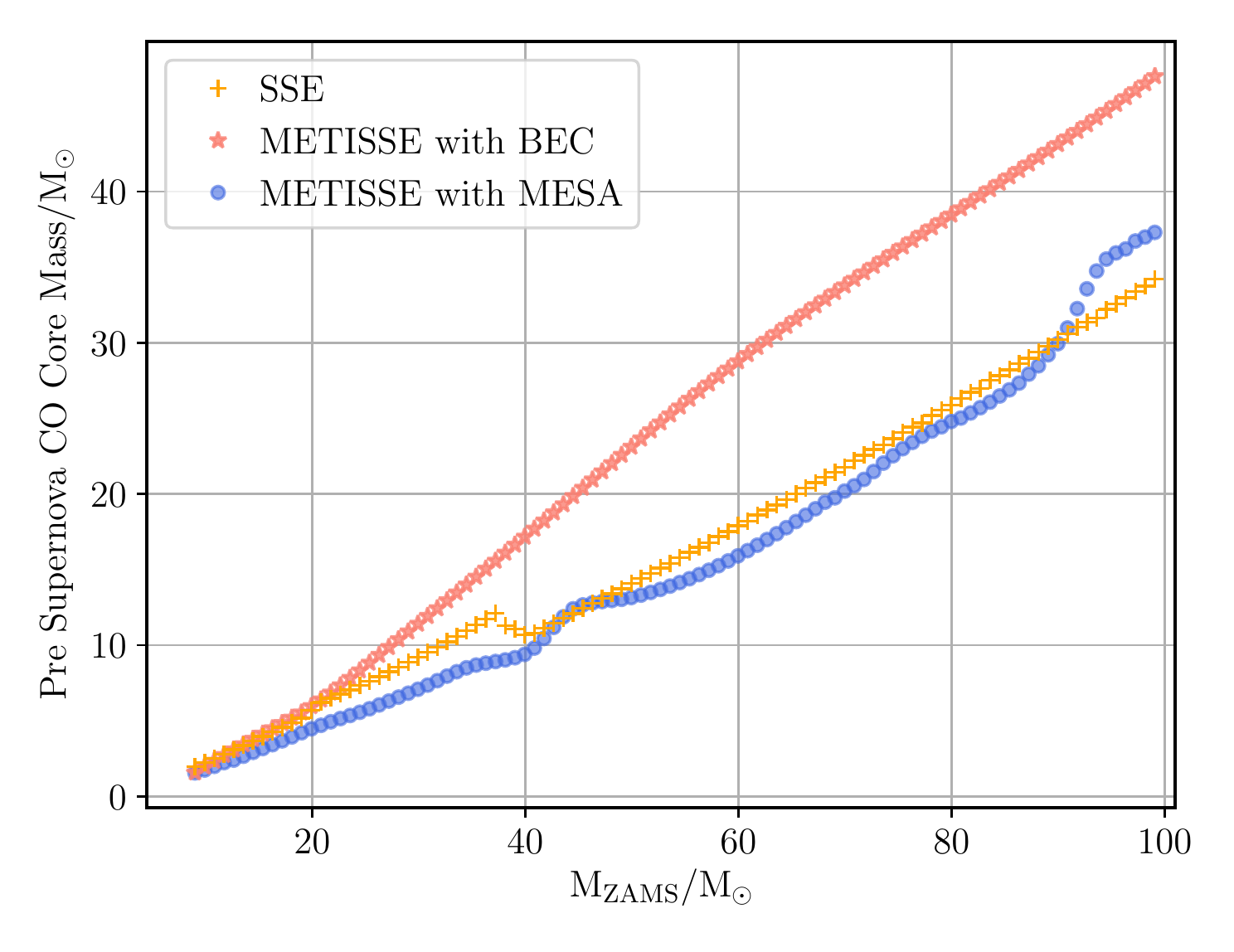}
    \\
    \end{tabular}
	\caption{Pre-supernova mass \textit{(left)} and final CO core mass \textit{(right)} of stars as a function of their ZAMS mass, symbols are the same as in Fig.~\ref{fig:rem}.}
	\label{fig:mass_presn}
\end{figure*}

In Fig.~\ref{fig:mass_presn} we plot the total mass and the core mass of stars (before supernova explosion) with respect to their initial mass as given by \textsc{SSE}, \textsc{METISSE} with \textsc{MESA} models and \textsc{METISSE} with \textsc{BEC} models.
The total masses for \textsc{MESA} and \textsc{BEC} tracks show only a small variation until 40\Msun{}. Beyond 40\Msun{}, stars evolved with \textsc{MESA} start experiencing increased wind mass-loss rates owing to their proximity to the Eddington limit and hence end up with a lower mass.
Stars evolved with \textsc{BEC} in the 40 to 100\Msun{} range undergo envelope inflation as they encounter the Eddington limit inside the envelope. They experience owing due to the cool supergiant phase (see details in Section~\ref{subsec:disc_mass_loss_scheme}). 
None of the stars in the \textsc{BEC} models used here lose their envelope completely. 
Hence their remnant masses and maximal radii increase almost linearly in this region and are higher than those of \textsc{MESA} tracks. 

The extrapolation of stellar models in the \textsc{BEC} and the MLT++ method of \textsc{MESA} are numerical solutions employed to push forward the evolution of massive stars when they encounter the Eddington limit. A more accurate treatment of the super-Eddington limit in 1D stellar evolution codes is not available. In fact, recent 3D simulations show that the 1D stellar evolution codes might not be modelling these envelopes accurately at all \citep{2015ApJ...813...74J,2018Natur.561..498J}. Note that \textsc{METISSE} provides enough flexibility that, if new stellar models with an updated treatment of Eddington limit proximity are published in the future, it will be straightforward to use them with \textsc{METISSE}.

\subsection{Mass loss schemes}
\label{subsec:disc_mass_loss_scheme}

Depending on the mass, the chemical composition and the evolutionary phase of a star, mass loss through stellar winds can have a considerable effect on the evolution of stars. For massive stars, wind mass loss and its role in stellar evolution is particularly important \citep[see e.g.][]{2014ARA&A..52..487S, 2017A&A...603A.118R}. 

Proximity to the Eddington limit on the stellar surface can also lead to departure from hydrostatic equilibrium and possible turbulence and mass outflows that exhibit as enhanced stellar winds \citep{1994PASP..106.1025H,2004ApJ...616..525O}. Although very massive stars (in the form of luminous blue variables) have been observed to undergo such high wind mass loss episodes, the presence of super-Eddington winds and their exact contribution is unconfirmed and remains a debated topic in the literature \citep{1997ASPC..120...83L,2017RSPTA.37560268S}. 

The \textsc{MESA} models used here are computed with mass-loss rates from \citet{2001A&A...369..574V} for ${\rm T_{eff} > 10\,000\,K}$ and \citet*{1988A&AS...72..259D} for ${\rm T_{eff} < 10\,000\,K}$. In addition, they include a contribution to mass loss by super-Eddington winds. We find that the super-Eddington mass-loss rate calculated in the default \textsc{MESA} can be extremely high (about $10^{-2}$\Msun{}\,yr$^{-1}$). Hence we scale down the super-Eddington wind mass loss by a factor of 10 and only apply it whenever the surface luminosity exceeds 1.1~times the mean Eddington luminosity (mass-weighted average of $L_{\rm {Edd}}$, see Equation~\ref{eq:ledd}, from the surface up to the region with optical depth of 100). 
Additionally, the maximum mass-loss rate we allow is capped to $1.4 \times 10^{-4}$\Msun{}\,yr$^{-1}$ following \citet{2010ApJ...714.1217B}. 

The \textsc{BEC} tracks also include mass-loss rates of \citet{2001A&A...369..574V} for ${\rm T_{eff} > 22\,000\,K}$. Below this, the mass-loss rate of \citet{1990A&A...231..134N} is applied whenever it exceeds the rate of \citet{2001A&A...369..574V}.
Although an enhancement of the mass loss due to rotation is an option in \textsc{BEC} as per \citet{2005A&A...435..967Y}, the models here are slowly rotating (at $100$\,km s$^{-1}$) and thus the rotational enhancement of mass loss (which becomes important when the star rotates close to the Keplerian critical rotational rate) does not contribute significantly. 

\begin{figure}
  \includegraphics[width=\columnwidth]{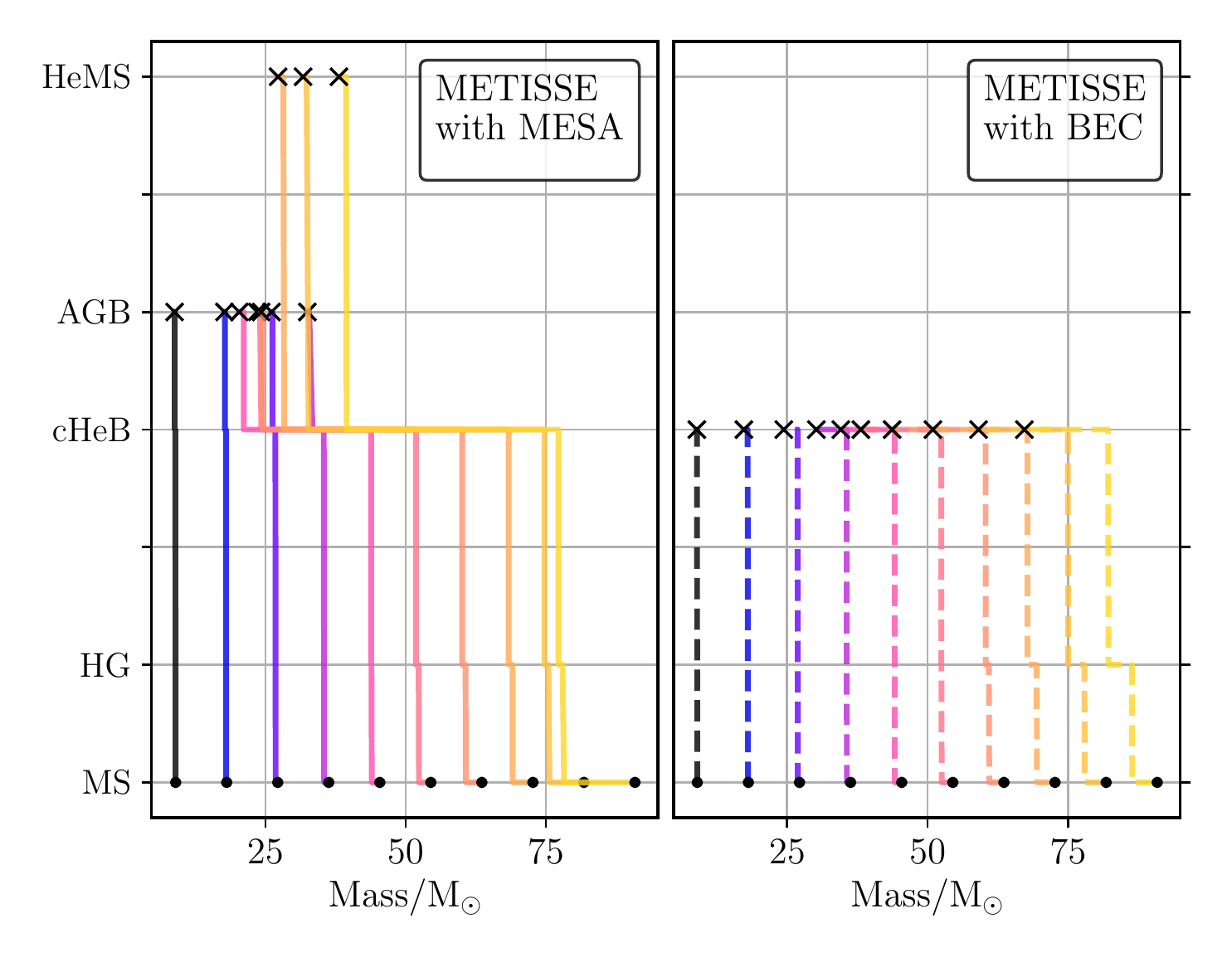}
  \caption{Mass of the star at different evolutionary phases calculated by \textsc{METISSE} with \textsc{MESA} (left panel, solid lines) tracks and \textsc{BEC} tracks (right panel, dashed lines). For each star, the dot represents the initial mass and the cross represents the pre-supernova mass. For explanation see Section~\ref{subsec:disc_mass_loss_scheme}}
  \label{fig:mass_phase}
\end{figure}

To examine the effect of the above schemes on the results obtained in Section~\ref{sec:compare_stellar_models} we plot the total mass of stars during different evolutionary phases in Fig.~\ref{fig:mass_phase}. 
We find that for both \textsc{MESA} and \textsc{BEC}, most of the mass loss happens towards the end of the core hydrogen burning (MS) and core helium burning (cHeB) phases. 
Towards the end of the main sequence, when \textsc{BEC} models become H-burning cool supergiants, the major contribution to mass loss comes from the supergiant mass-loss rates of \citet{1990A&A...231..134N}. 
Stars evolved with \textsc{MESA}, on the other hand, either experience mass loss according to \citet{2001A&A...369..574V} or through super Eddington winds, depending on whether their surface luminosity exceeds the Eddington limit by 10 per cent.

As shown in Fig.~\ref{fig:He_fraction}, \textsc{MESA} models at 20\Msun{} and above burn helium at different effective temperatures to those from the Bonn code, and therefore experience a different kind of mass-loss treatment. 
The massive stars in the \textsc{BEC} tracks experience mass-loss rates of \citet{1990A&A...231..134N} during core helium burning due to their low effective temperatures but now at high luminosity. Hence, they lose more mass during this phase than towards the end of the main sequence. 
The \textsc{MESA} tracks up to 40\Msun{} demonstrate moderate mass loss during cHeB because the models continue their slow transition from \citet{2001A&A...369..574V} to \citet{1988A&AS...72..259D} mass-loss rates.
More massive stars with \textsc{MESA}, those experiencing super-Eddington winds, can lose their envelopes completely and become naked helium stars when this major mass-loss episode kicks in during cHeB. The remainder of the evolution of such stars is performed with fitting formulae for helium star models from \textsc{SSE} and the mass-loss scheme of \citet{1995A&A...299..151H} is applied
(see Section~\ref{subsec:stellar_phases} for details).
We find that \textsc{MESA} stars do not spend much time in this phase and, as shown in Fig.~\ref{fig:mass_phase}, hardly lose any mass.

For \textsc{SSE} tracks, mass-loss rates have been calculated by \citet{2010ApJ...714.1217B}. Stars above 38\Msun{} experience mass loss at $1.5 \times 10^{-4}$\Msun{}\,yr$^{-1}$ whenever the surface luminosity $(L)$ exceeds $10^{5}L$\sun{} and radius $(R)$ satisfies $10^{-5} R L^{0.5}>1.0 R\sun L\sun^{0.5}$ \citep[see equation 8 of][]{2010ApJ...714.1217B}, and end up with lower remnant masses, similar to models evolved with \textsc{MESA} with \textsc{METISSE}. 

Chemical composition also plays a key role in determining mass-loss rates. Stars with higher metal content have higher opacities and therefore have higher mass-loss rates \citep{2001A&A...369..574V, 2015IAUS..307...25P}.  
Following \citet{2001A&A...369..574V}, a metallicity dependence of Z$^{0.86}$ is included into the treatment of mass loss in all models. The stellar models here have approximately the same Z, that is, nearly one tenth of solar as per \citet{2009ARA&A..47..481A}. The initial metallicity of the \textsc{MESA} models is $Z = 0.00142$ with element ratios scaled down from solar composition. \textsc{BEC} models are computed with $Z = 0.00105$ and have chemical composition scaled by a factor of 2 down from that of the Small Magellanic Cloud (SMC). 
While differences in the abundances of individual metal elements also influence the opacity and energy transport rates, the main contributors to the winds of massive stars are iron-like elements \citep*{2000A&AS..141...23P}. As shown in fig.~1 of \citet{Szecsi:2015}, except for carbon and nitrogen, SMC abundances are proportional to those of Solar and the contribution of these two elements to line driving (and thus to mass loss) is relatively minor.

\subsection{Convection and overshooting}
\label{subsec:disc_conv}

Massive stars have convective cores owing to a steep temperature gradient in the interior. These cores can overshoot beyond convective boundaries into non-convective regions due to finite particle velocities and cause enhanced mixing of elements inside stars \citep{1963ApJ...138..297B,1973ApJ...184..191S,1975A&A....40..303M}. Thus the location of convective boundaries is important to determine the evolution of massive stellar cores and the lifetimes of different evolutionary phases of a star \citep{2012ARA&A..50..107L}. 

Convection and overshooting are complex 3D processes, although in 1D stellar evolution codes they are treated using the mixing length theory \citep[MLT;][]{1958ZA.....46..108B} or some modified version of it.
Convection is modelled in terms of the mixing length parameter ${\rm \alpha_{MLT}}$ over a region determined by the Ledoux or Schwarzchild criteria \citep{1990sse..book.....K}.  Overshooting can be modelled with the step overshoot prescription where the convective boundary is simply extended by a fraction of the pressure scale height, given by the parameter ${\rm \delta_{ov}}$ \citep{1963ApJ...138..297B, 1975ApJ...198..407S}.

The \textsc{MESA} models in this work use convection parameters calibrated to the Sun and a modified version of MLT by \citet*{1965ApJ...142..841H} with ${\rm \alpha_{MLT}}= 1.82$ \citep{2016ApJ...823..102C}. Convective-overshoot is assumed to vanish exponentially outside the convective region with ${\delta_\mathrm{ov}} = 0.016$  
\citep[][]{2010ApJ...718.1378M, 2016ApJ...823..102C}. This is roughly equivalent to $\delta_\mathrm{ov} = 0.2$ in the step overshooting model.

On the other hand, \textsc{BEC} models have used standard MLT \citep{1958ZA.....46..108B} with ${\rm \alpha_{MLT}} = 1.5$ and step overshoot with ${\delta_\mathrm{ov}} = 0.335$ \citep{1991A&A...252..669L, Brott:2011}. 
The overshoot values have been calibrated to massive stars observed with the VLT-FLAMES survey \citep{2008A&A...479..541H}.
Both the \textsc{MESA} and \textsc{BEC} models have the Ledoux criterion to determine convective boundaries.

Stars evolved with \textsc{MESA} have larger convective efficiency owing to the greater ${\rm \alpha_{MLT}}$ which decays exponentially
outside the fiducial convective region while the larger
$\delta_\mathrm{ov}$ in the \textsc{BEC} models means a more extended region for
convection. However the smaller value of ${\rm \alpha_{MLT}}$ in \textsc{BEC} reduces mixing efficiency, leading to less mixing overall and hence shorter MS lifetimes compared to \textsc{MESA} models (Fig.~\ref{fig:ms_lifetime}).

In \textsc{SSE}, the formulae for determining main-sequence lifetimes were calculated with models of \citet{1998MNRAS.298..525P}, where the authors adopted standard MLT with ${\rm \alpha_{MLT}} = 2.0$ while overshooting was modelled through a modification of the Schwarzchild criterion.
Their overshooting coefficient $(\delta_\mathrm{ov} = 0.12)$ approximates to ${\delta_\mathrm{ov}}\sim0.4$ in the step overshooting prescription for the most massive stars ($\sim$50\Msun{}) in the set. Due to the high value of the overshooting parameter, combined with the efficient mixing length parameter in the \citet{1998MNRAS.298..525P} tracks, \textsc{SSE} predicts even longer MS lifetimes for stars compared to the \textsc{MESA} and \textsc{BEC} models (Fig.~\ref{fig:ms_lifetime}).

Mixing processes are not very well constrained for massive stars \citep{2019A&A...625A.132S}. In particular, the convection and overshooting parameters are sensitive to quantities like opacities \citep{1991ApJ...381L..67S} and the Solar abundance scale \citep{2010ApJ...718.1378M}. Processes such as semiconvection \citep*{1983A&A...126..207L} and rotational mixing \citep{Heger:2000a} also contribute significantly to the evolution of such stars. There are ongoing efforts to improve constraints on the mixing parameters with 3D hydrodynamic simulations (\citealp{2014MNRAS.445.4366T}, \citealp*{2015A&A...573A..89M}) and asteroseismic measurements \citep{2010Ap&SS.328..227N}. 
\textsc{METISSE} can be useful in the future to study the effects of varying different mixing parameters and comparing the outcome to observed populations of massive stars. This is a major advantage over an \textsc{SSE} style code in that any piece of physics in the stellar models can be changed without having to find a whole new set of fitting formulae.

\section{Caveats and future work}
\label{sec:caveats_and_future_work}

Interpolation has the advantage of being fast, robust and able to utilize different sets of stellar evolution models with ease. As with any other method, it has some limitations as well. We discuss some of these in this section and how \textsc{METISSE} aims to address them.

\subsection{Quality and completeness of input stellar tracks}

Results produced by \textsc{METISSE} are a direct reflection of the quality of the input stellar models. Fine details in the input models
can be reproduced but so can the flaws.
For the calculation of stellar tracks up to their respective remnant phases, input models should at least be evolved until the formation of a CO core, because the CO core mass is needed to calculate remnant properties \citep{2008ApJS..174..223B,2012ApJ...749...91F}. To avoid propagation of inaccuracies of the stellar models in the results obtained, input tracks need to be checked thoroughly for flaws and incompleteness. 

The set of input stellar models should also be dense, particularly near mass cutoffs, to ensure accuracy of the interpolation. 
If some tracks are incomplete due to convergence issues, \textsc{METISSE} can attempt to calculate the missing phase as described in Section~\ref{subsec:incomplete_track}. However, it fails if many tracks are incomplete over a small mass range or if the set of input models is too sparse.

\subsection{Mass and metallicity limits}

Currently in \textsc{METISSE} stellar tracks can only be interpolated for the same metallicity as the input models. Although interpolation between tracks of different metallicity could be implemented, the interpolated track might not be a good approximation of a detailed track of the same mass and metallicity unless these two metallicities are sufficiently close. Even then, tracks for interpolation will have to be carefully selected because the occurrence of evolutionary features in a track also depends on the metallicity.

Interpolation in \textsc{METISSE} is also bounded between the highest and lowest mass track present in the set of input models.
Extrapolation can lead to spurious results if the tracks used for the extrapolation are sparsely distributed. 
Hence, we do not extrapolate beyond the maximum mass track of the input set in \textsc{METISSE}. 
However, we do extrapolate a new track from higher masses if an input mass falls between a critical mass (cf. Section~\ref{subsec:Z_par}) and the initial mass of the next track. Because the density of stellar tracks where these mass cut-offs occur is usually high, the tracks obtained are a suitable approximation to the
evolution of such stars. We are currently working on a new version of \textsc{METISSE} to further limit the reliance on extrapolation
near mass cutoffs.

\subsection{Information about stellar structure}

Stars in binary systems can transfer mass on to each other if they expand beyond their Roche-lobe radii. If the mass transfer is significant, it can affect the structure and the evolution of the member stars. It can, for example, affect the structure of the core and the burning shells \citep{2017A&A...603A.118R} which can be crucial to determine the type of remnant formed by the star.
Therefore, details of the stellar interior are needed to accurately compute properties of stars in response to mass transfer in a binary system.

Creation and storage of large sets of stellar evolution models and the use of them in conjunction with detailed codes for binaries as done by the \textsc{Binary Population and Spectral Synthesis} \citep[BPASS;][]{2016MNRAS.462.3302E, 2017PASA...34...58E} project and the \textsc{Brussels code} \citep{2004NewAR..48..861D} is another way to account for the stellar structure in response to mass transfer in binaries. However, it requires expertise to maintain and run such models.

We intend to apply \textsc{METISSE} in binary population studies in the future. While \textsc{METISSE} cannot compute changes to the internal structure of a star in response to mass transfer, it can easily interpolate between any stellar structure parameters provided by the detailed models, thus extending more widely than the main parameters of total mass, core mass, luminosity and radius. 
 
Examples include the mass of the convective envelope (important for mass transfer), the moment of inertia \citep[important for tidal evolution, c.f.][]{2013ApJ...764..166D} and the envelope binding energy \citep*[important for common envelope evolution, c.f.][]{Loveridge:2011ApJ,2010ApJ...716..114X}.

\section{Conclusions}
\label{sec:conclusions}

We have presented our new code \textsc{METISSE} and its capabilities as a standalone synthetic stellar evolution code. \textsc{METISSE} can simulate stars from the ZAMS to the end of the full range of stellar remnant phases, including naked helium star phases. We find that \textsc{METISSE} better reproduces stellar tracks than the \textsc{SSE} fitting formulae with the same input data. \textsc{METISSE} is similar in performance to \textsc{SSE} with the added advantage that it can be easily used with different sets of stellar evolution tracks. 

Massive stars are the progenitors of compact objects, neutron stars and black holes, whose merging result in the emission of gravitational waves observable by LIGO/Virgo \citep{2016PhRvL.116f1102A}. 
We have used \textsc{METISSE} to demonstrate that uncertainties in modelling the evolution of massive stars, such as their radiation dominated envelopes, can have a remarkable influence on their evolution. 
Such uncertainties can impact the radial expansion of stars and the properties of stellar remnants, which can subsequently change the interactions in binary and star cluster environments. 
Therefore, the ability to accurately predict stellar remnant masses is crucial when attempting to account for present day observations of compact object populations. 

Surveys dedicated to the study of massive stars \citep{2011A&A...530A.108E,2011AN....332..232K, 2020arXiv200202690V} have advanced our understanding of these stars. 
In the coming years, instruments such as the \textit{James Webb Space Telescope} \citep[JWST:][]{2003IAUJD...8E..10G}, the \textit{Giant Magellan Telescope} \citep[GMT:][]{2014tmt..confE..61M},
the \textit{Large Synoptic Survey Telescope} \citep[LSST:][]{2009arXiv0912.0201L} and the \textit{Laser Interferometer Space Antenna} \citep[LISA:][]{2017arXiv170200786A} will further boost our knowledge of stars and stellar systems. As the data from newer observations becomes available,
and the stellar structure and evolution codes become better at modelling stellar phenomena, both in 1D and 3D, we will be able to include the updated stellar models in our population synthesis codes through \textsc{METISSE}.

Since \textsc{METISSE} has been written in the same variable and file structure as \textsc{SSE}, it will be easy to include it in population synthesis codes as an alternative to \textsc{SSE} \citep{Hurley2000}.
In the future we plan to publicly release \textsc{METISSE} as well as integrate it with the binary population synthesis codes \textsc{BSE} \citep*{2002MNRAS.329..897H} and \textsc{COMPAS} \citep{2017NatCo...814906S, 2018MNRAS.481.4009V} and the star cluster modelling code \textsc{NBODY6} \citep{aarseth_2003}. Using \textsc{METISSE} will not only help us to include up to date treatments of stellar evolution in population synthesis codes but will also enable us to study the role of different stellar evolution parameters on the evolution of stellar systems and make predictions to pave the way for the new missions.

\section*{Acknowledgements}
We thank Carolyn Doherty, Ilya Mandel, Alejandro Vigna-G\'{o}mez, Coen Neijssel, Jan Eldridge, Simon Campbell, Alex Kemp and Isobel-Romero Shaw for useful comments and suggestions. 
We also thank Aaron Dotter for making ISO open source.
We are also grateful to the referee, Chris Tout, for helpful comments.
DSz has been supported by the Alexander von Humboldt Foundation. 
PA, JH and SS acknowledge the support by the Australian Research Council Centre of Excellence for Gravitational Wave Discovery (OzGrav), through project number CE170100004. This work made use of the OzSTAR high performance computer at Swinburne University of Technology. OzSTAR is funded by Swinburne University of Technology and the National Collaborative Research Infrastructure Strategy (NCRIS).

%%%%%%%%%%%%%%%%%%%%%%%%%%%%%%%%%%%%%%%%%%%%%%%%%%



\bibliographystyle{mnras}
\bibliography{bibliography.bib} % if your bibtex file is called example.bib

%%%%%%%%%%%%%%%%%%%%%%%%%%%%%%%%%%%%%%%%%%%%%%%%%%

%%%%%%%%%%%%%%%%% APPENDICES %%%%%%%%%%%%%%%%%%%%%

\appendix

\section{extra details on the \textsc{METISSE} methodology}

\subsection{Z-parameters and Mass cutoffs}
\label{subsec:Z_par}
Searching within the set of stellar tracks to find neighbouring mass tracks for interpolation may seem straight forward but there is a catch to it. The tracks of neighbouring masses are usually similar in properties but certain features occur only in a range of masses and not in others. Interpolation between these can result in incorrect tracks.

Hence, we define five critical masses similar to those defined by \citet{1998MNRAS.298..525P}. These critical masses or Z-parameters are fixed for a given metallicity, and serve as the lower limits above which certain physical properties start to appear for stellar tracks. They are

\begin{itemize}[leftmargin=.3in]
    \item $M_\mathrm{hook}$ -- Mass above which the hook feature starts to appear on the MS, 
    \item $M_\mathrm{HeF}$ -- Mass above which He ignition occurs non-degenerately in the core,
    \item $M_\mathrm{FGB}$ -- Mass above which He ignition occurs on the HG,
    \item $M_\mathrm{up}$ -- Mass above which off-centre C/O ignition can occur non-degenerately in the core and
    \item $M_\mathrm{ec}$ -- Mass above which a star avoids electron captures on neon and proceeds to form an iron core. 
     
\end{itemize}

Some Z-parameters correspond to the behaviour of core properties and are useful to determine the type of remnant a star will become. The locations at which these critical masses occur in the set are stored as an array of mass cutoffs. For any input mass, only tracks whose initial masses are located within the mass cutoffs are used for the interpolation in mass. 
If the input mass falls between a critical mass and the initial mass of the next track then its track is extrapolated from the higher mass tracks. 

Use of these critical masses not only helps avoid interpolation between dissimilar tracks but also narrows down the range to search for the nearest mass track, thus saving computation time. Z-parameters and mass cutoffs are automatically located by \textsc{METISSE} for any input set of stellar models.
If the automatic location method fails to provide a correct value, then \textsc{METISSE} has the option of using Z-parameters supplied by the user. 

\subsection{Stellar Remnants}
\label{subsec:stellar_remnants}

In \textsc{METISSE}, when the star reaches the end of its nuclear burning life either after the AGB or by satisfying equation~\ref{eqn:mcmax}, it becomes a remnant. The type of remnant formed depends on whether or not the core of the star is able to ignite carbon, and if the ignition leads to the formation of an iron core, which can gravitationally collapse in a supernova. Hence, in \textsc{METISSE} we utilize the corresponding critical mass cutoffs $M_\mathrm{up}$ and $M_\mathrm{ec}$ 
(as defined above in Sec.~\ref{subsec:Z_par}) 
in the decision-making. 

The type and mass of the remnant formed by a star can then be determined by comparison of the He core mass of the star at the base of the AGB $(M_\mathrm{c,BAGB})$ to the core masses at $M_\mathrm{up}$ $(M_\mathrm{up,core})$ and $M_\mathrm{ec}$ $(M_\mathrm{ec,core})$ as described below while other properties (e.g., luminosity and radius) are calculated with \textsc{SSE} formulae \citep[see section 6.2 of][]{Hurley2000}. The outcomes are as follows, 

\begin{enumerate}
    \item \textbf{White Dwarf:}
    if the final CO mass of the star is less than the Chandrasekhar mass $(M_\mathrm{ch})$, it can either become  
    a carbon-oxygen white dwarf (CO-WD) if ${M_\mathrm{c,BAGB} < M_\mathrm{up,core}}$ or an oxygen-neon white dwarf (ONe-WD) if $ {M_\mathrm{c,BAGB} \geq M_\mathrm{up,core}}$. 
    The mass of the white dwarf is taken to be the same as the final CO core mass of the star.
    
    \item \textbf{Neutron Star or Black Hole:}
    if the final CO core mass of a star exceeds $M_\mathrm{ch}$, it is assumed to explode in a supernova. If ${M_\mathrm{c,BAGB} < M_\mathrm{up,core}}$, then the carbon ignites under degenerate conditions and the star leaves behind no remnant. On the other hand, if ${M_\mathrm{c,BAGB} \geq M_\mathrm{ec,core}}$ the star undergoes core-collapse to form either a neutron star or black hole.
    The type and mass of the resulting compact remnant can be calculated from one of the following prescriptions: (a) \citet*{2002ApJ...572..407B} (b) \citet{2004MNRAS.353...87E} (c) \citet{2008ApJS..174..223B}. In between the two limits the star is assumed to explode as an electron-capture supernova and form a neutron star of 1.26\Msun{}.
        
\end{enumerate}

\subsection{Calculation of missing phases}
\label{subsec:incomplete_track}

With the EEP based format, one can define a limit on the number of data points depending on how many phases a particular track has. This can be different for stars that undergo C burning to become a neutron star or a black hole to that of stars that form a white dwarf. Due to numerical and convergence issues there can be incomplete tracks present in the input set of models. In such cases, even when a single track used for interpolation has insufficient data points, the interpolated track is also rendered incomplete. 

Hence, in \textsc{METISSE} we check if, after interpolation, the track has a certain minimum required number of points. By default, this minimum is the TPAGB for low-mass stars and end of C burning for high-mass stars. Both limits can be changed by the user for different sets of stellar models. If the track is incomplete, \textsc{METISSE} searches in the input set (within mass cutoffs) for complete tracks closest to the input mass and interpolates the remaining track from there.  
The method works only if there are at least two complete tracks within the mass cutoff and there are no large mass gaps in the input set.

%%%%%%%%%%%%%%%%%%%%%%%%%%%%%%%%%%%%%%%%%%%%%%%%%%

% Don't change these lines
\bsp	% typesetting comment
\label{lastpage}
\end{document}